\documentclass[prb,superscriptaddress,twocolumn,showpacs]{revtex4} 
\usepackage{graphicx}
\usepackage{amsmath,amssymb,bm} 
\usepackage{textcomp} 
\usepackage{xspace} 
\usepackage{soul,hyperref}
\usepackage[usenames,dvipsnames]{color}
\usepackage{subfigure}
\usepackage{pstool}

\newcommand{\tmyag}{Tm\ensuremath{^{3+}}:YAG\xspace}
\newcommand{\tmtrans}{{\ensuremath{^3}H\ensuremath{_6(0)\rightarrow ^3}H\ensuremath{_4(0)}}\xspace}
\newcommand{\iso}[1]{\ensuremath{{^{#1}}}} 
\newcommand{\tplus}{\ensuremath{^{3+}}\xspace}

\begin{document}

\title{Optical measurement of heteronuclear cross-relaxation interactions in Tm:YAG}

\author{R. L. Ahlefeldt}
\affiliation{Laboratoire Aim\'e Cotton, CNRS-UPR 3321, Univ. Paris-Sud, B\^at. 505, 91405 Orsay cedex, France}

\author{M. F. Pascual-Winter}
\affiliation{Centro At\'omico Bariloche and Instituto Balseiro, C.N.E.A., 8400 S. C. de Bariloche, R. N., Argentina}

\author{A. Louchet-Chauvet}
\affiliation{Laboratoire Aim\'e Cotton, CNRS-UPR 3321, Univ. Paris-Sud, B\^at. 505, 91405 Orsay cedex, France}

\author{T. Chaneli\`{e}re}
\affiliation{Laboratoire Aim\'e Cotton, CNRS-UPR 3321, Univ. Paris-Sud, B\^at. 505, 91405 Orsay cedex, France}

\author{J.-L Le Gou\"{e}t}
\affiliation{Laboratoire Aim\'e Cotton, CNRS-UPR 3321, Univ. Paris-Sud, B\^at. 505, 91405 Orsay cedex, France}

\date{\today}
\begin{abstract}
We investigate cross-relaxation interactions between Tm and Al in \tmyag using two optical methods: spectral holeburning and stimulated echoes. These interactions lead to a substantial reduction in the hyperfine lifetime at magnetic fields that bring the Tm hyperfine transition into resonance with an Al transition.  We develop models for the measured echo decay curves and holeburning spectra near a resonance, which are used to show that the Tm-Al interaction has a resonance width of $12\pm 3$~kHz and reduces the hyperfine lifetime at resonance to $0.5\pm 0.3$~ms. 
\end{abstract}


\maketitle

\section{Introduction}
Rare earth ions in crystals have the unique property of maintaining long coherence times at high spatial density. This property lies at the root of the sustained interest in these materials for information processing applications in both the classical and quantum domains. When combined with large inhomogeneous linewidths, long coherence times allow classical processing with very large time-to-bandwidth products, and  applications of rare earth materials to RADAR, electronic warfare, and astrophysics have been pursued actively for the last two decades (Ref \onlinecite{babbitt14} and references therein). These applications favor materials such as Tm\tplus:YAG (Y$_{3}$Al$_{5}$O$_{12}$) and Er\tplus:Y$_2$SiO$_5$ as they have a one-to-one correspondence between  the signal frequency  and  the storage ion transition frequency.

For quantum applications, the long coherence times of the easily accessed rare earth optical transitions already make them good candidates, but the possibility of transferring optical coherence to much longer lived spin coherence makes these materials particularly attractive. The non-Kramers ions Pr, Eu, and Tm in low symmetry sites, such as  in YAG and Y$_2$SiO$_5$, are particularly useful, as they have singlet electronic levels that are highly insensitive to the crystalline environment, a property that favors the existence of very long coherence times. The quantum applications of rare earth ions are reviewed in Refs \onlinecite{thiel11} and \onlinecite{tittel10}.

Optimising performance for  these applications requires a detailed understanding of the way the rare earth ion interacts with its crystalline environment. Ion-ion interactions affect properties such as the optical density, inhomogeneous linewidth, and level lifetimes, and are often the dominant source of decoherence.  For example, in Er\tplus:Y$_2$SiO$_5$, B\"ottger et al. showed that  interactions between Er ions dominate optical dephasing and that this effect can be avoided by using low concentration crystals in high magnetic fields, resulting in a coherence time of 2~ms, the longest optical coherence time observed in a solid \cite{bottger06}. For hyperfine transitions of non-Kramers ions, meanwhile, dephasing usually comes from the magnetic fluctuations due to flipping lattice spins, and suppressing this effect contributed to the longest hyperfine coherence times measured, of 6 hours \cite{zhong15}.

Despite those unquestionable successes, the dynamics of the host crystal, a dominant source of decoherence, have still to be better understood. In the present work we intend to gain information on the dynamics of lattice ions through their resonant interaction with optically active, low-concentration, impurities. In this paper, we use the term ``resonant interaction'' to refer specifically to spin polarization transfer between the interacting ions.

Heteronuclear resonant interactions, arising from the magnetic dipole-dipole coupling of unlike nuclei, are well known in NMR spectroscopy. They are used in a variety of common techniques, including nuclear Overhauser spectroscopy\cite{anderson62}, cross polarisation spectroscopy \cite{hartmann62}, and adiabatic demagnetisation  \cite{anderson62a}, for detecting low concentration or low magnetic moment species by transferring spin polarization from a high concentration, high moment species. The situation we consider is just the opposite. The high sensitivity of optical detection enables us to directly monitor scarce impurity ions. The lattice dynamics are little affected by those ions, but can be seen in their time evolution. We take advantage of resonance to precisely select the lattice ions that are probed by the impurities.

Techniques based on resonant interactions between unlike nuclei have very seldom been used in optical spectroscopy, particularly in the high resolution spectroscopy used to study rare earth materials. We are only aware previous work on one material, Pr:LaF$_3$ \cite{wald92,otto86,lukac89}. In this material, strong  resonant interactions between Pr and F occur.  These were used to prove the existence of a frozen core of fluorine spins \cite{wald92}. 

In this paper we investigate resonant interactions between Tm and its Al neighbors in \tmyag, a material considered for both classical and quantum information processing applications \cite{macfarlane93, guillot-noel05, babbitt14, heshami12}.  These interactions cause sharp decreases in the hyperfine state lifetime when a magnetic field brings the two ions into resonance. This occurs for fields $< 10$ mT, a region commonly used for information processing applications. High temperature NMR investigation of the resonant exchange of excitation between Tm and Al in the isomorphic crystal TmAlG has been undertaken in the past \cite{trontelj73}. It was found that the resonant interaction with the abundant Tm ions can facilitate spin-lattice relaxation of $^{27}$Al. In this paper, we work with  a low concentration of Tm, precisely so that the Al dynamics are not altered.

Our investigation is based on two common optical spectroscopy methods, spectral holeburning and stimulated echoes. We start by giving examples of how ion-ion resonant interactions affect the holeburning spectrum in Section \ref{sec:ext}. In Section \ref{sec:lifetimes}, we characterize the relaxation behavior of the system away from resonance using stimulated echoes. We then use this information to develop a simple model for the echo decay at resonance. In Section \ref{sec:struct}, we extend this model to the more complicated case of the holeburning structure. In Section \ref{sec:zeeman}, we include the Al Zeeman interaction into this model, and in Section \ref{sec:anticomp}, compare the model to the experimental holeburning data. Finally, in Section \ref{sec:discussion} we briefly discuss the resonant interaction mechanism, and the possibility of directly modelling the dynamic transfer of the Tm excitation to the spin bath.

\section{Hole burning extinction due to resonant interactions} \label{sec:ext}

\begin{figure}
	\centering
	\includegraphics[trim = 0 0 8 10,clip]{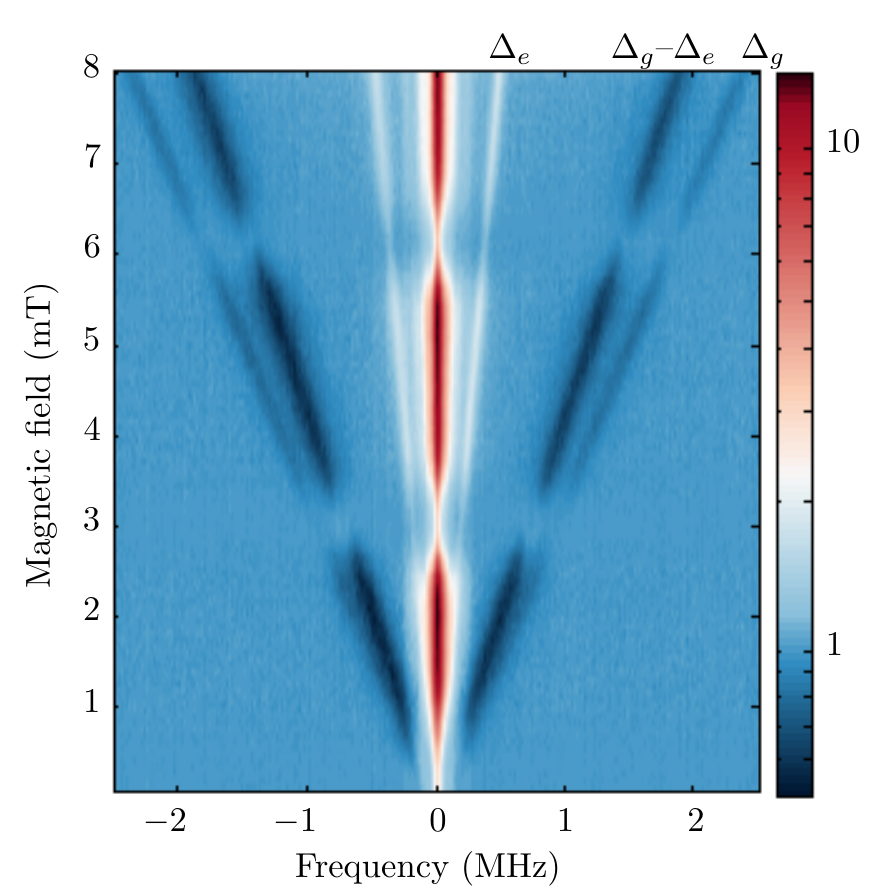}
	\caption{\label{fig:ext001} (Color online) Holeburning  transmission spectrum in \tmyag as a function of magnetic field along [001]. The color scale is logarithmic. One sidehole, and two antiholes are visible on each side of the central hole. These are labelled by their splitting from the central hole. Near complete extinction of the spectral hole occurs at 3.1~mT and 6.2~mT. Each horizontal spectrum is an average of eight shots. Between each shot the magnetic field was switched off, or to the nearest extinction point, to prevent accumulation of the spectral hole. }
\end{figure}

Tm has a single stable isotope, \iso{169}Tm, with a nuclear spin of $I = \frac{1}{2}$. In \tmyag, the Tm ion occupies the single Y site of D$_2$ symmetry. It has an optical transition from the $^3$H$_6$(0) ground state to the  $^3$H$_4$(0) excited state at 793 nm. In both the optical ground and excited states, the mixing of the electronic wavefunctions by the crystal field leads to a large and extremely anisotropic enhanced nuclear Zeeman tensor \cite{guillot-noel05, deseze06} of the order of 100~MHz/T.

Resonant ion-ion interactions have a particularly striking effect on the holeburning spectra of the \tmtrans transition of \tmyag. In zero magnetic field, \tmyag has a holeburning lifetime of $\approx 10$~ms due to storage in the $^3$F$_4$ metastable state. The lifetime increases to several minutes \cite{louchet07} when a magnetic field is applied as this splits the doubly degenerate ground state, allowing storage in the non-resonant hyperfine state. However, at certain magnetic fields, the hyperfine level lifetime plummets, leading to a decrease in the hole amplitude and the complete extinction of any antiholes. As we will show, these points of holeburning extinction occur when the magnetic field brings the Tm hyperfine transition into resonance with an Al transition.

  Extinction points can be seen in Figure~\ref{fig:ext001}, which shows the holeburning structure of the \tmtrans transition of \tmyag as a function of magnetic field magnitude. The experimental procedure used to generate the spectrum in Figure~\ref{fig:ext001} was as follows. A spectral hole was burnt with a weak, 100 ms pulse, and after a delay of 10~ms, a chirped pulse was used to read out the resulting holeburning structure. Each horizontal slice in the figure is the holeburning spectrum for a particular value of the applied magnetic field. The magnetic field was along the [001] direction of the crystal.  There are six magnetically inequivalent Tm sites in \tmyag, labelled as sites 1 to 6 (see Ref. \onlinecite{sun00}), and for a field orientation of [001] four sites (sites 3 to 6) are equivalent.  To eliminate the other two sites, sites 1 and 2, from the hole burning spectrum, the laser beam direction was chosen along [$\bar{1}10$] and the polarization along [001]. The holeburning spectrum, therefore, shows the structure of a single class of ions: one sidehole at $\Delta_e$, the excited state splitting, as well as two antiholes at $\Delta_g$, the ground state splitting, and $\Delta_g-\Delta_e$. A third antihole is expected at  $\Delta_g+\Delta_e$, but for the magnetic field orientation chosen, this is too weak to be seen \cite{deseze06}.

The figure shows that when the magnetic field is increased from zero the central hole amplitude increases as the ground state hyperfine levels split and the hyperfine lifetime starts to rise. However, at two fields, 3.1~mT and 6.2~mT, the central hole amplitude suddenly drops, and the antiholes disappear, indicating that the hyperfine ground state lifetime has decreased to less than the metastable state lifetime. The lifetime is substantially shortened over a wide region, 0.6~mT, which corresponds to a range in hyperfine splitting of approximately 200~kHz.

\begin{figure}
	\centering
	\includegraphics{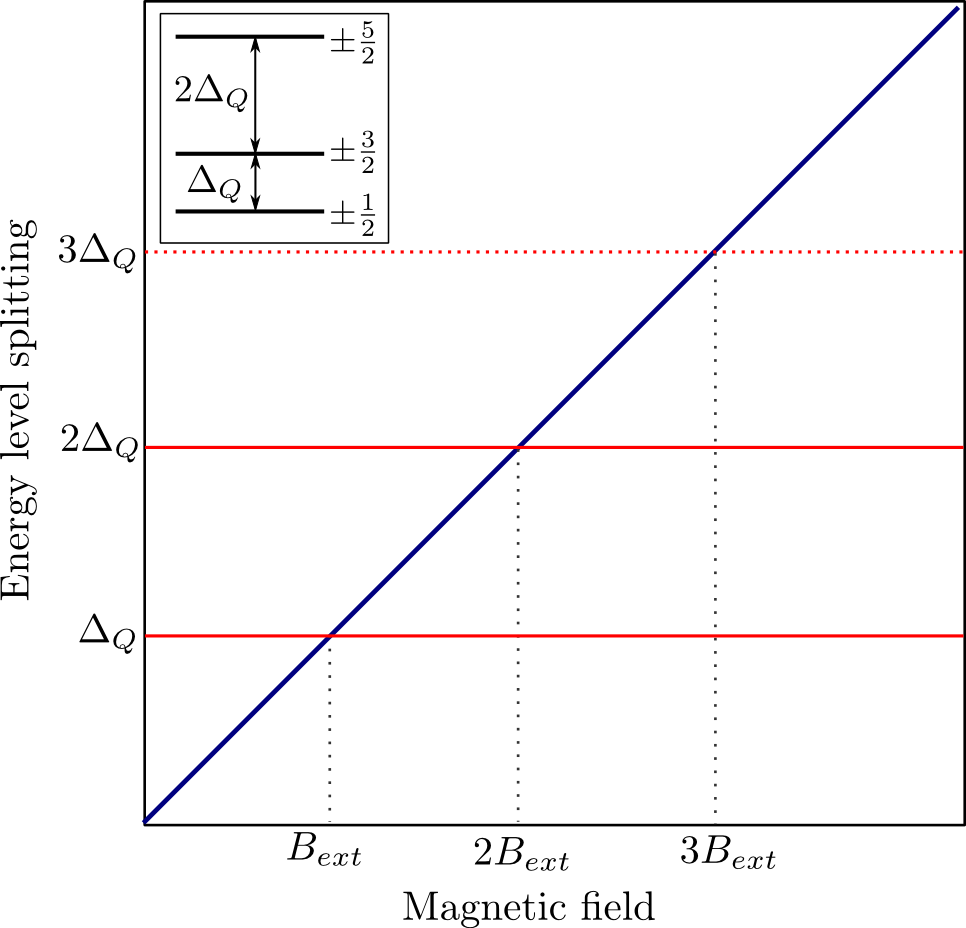}
	\caption{\label{fig:ext} (Color online) Holeburning extinction due to resonances between the Tm ground state splitting and the Al quadrupole splitting. The Al level structure is shown in the inset, and the horizontal red lines show the splittings between these levels. The dotted red horizontal line indicates an Al transition ($\pm 1/2 \rightarrow \pm 5/2$) that is extremely weak. Each time one of the Al splittings is in resonance with the Tm splitting (blue line), an extinction point can be expected. In Figure~\ref{fig:ext001} we observe the resonances at $B_{ext}$ and $2B_{ext}$. }
\end{figure} 
The hole extinction behavior can be explained by a resonant  cross relaxation between Tm and Al.  The interaction leads to a mutual spin flip-flop whereby the Tm excitation is transferred to the Al, filling in the Tm spectral hole. In Figure~\ref{fig:ext001}, we see two extinction points, with the second at twice the field of the first. This is clear evidence that the interacting species is Al.   Al has a nuclear spin of $I = 5/2$ and therefore a nuclear quadrupole moment. In the high symmetry Al sites of YAG, this tensor is cylindrically symmetric, giving rise to  three doubly degenerate hyperfine levels in zero field with splittings $\Delta_Q$ and $2\Delta_Q$. Al also has an isotropic nuclear Zeeman tensor, which lifts the double degeneracy of the three sets of levels in a magnetic field. However, because the nuclear magnetic moment of Al is more than an order of magnitude smaller than the Tm enhanced nuclear moment it has little effect on the position or behavior of the extinction points, and we will ignore it for now. It will be considered in detail in Section \ref{sec:zeeman}.

With this assumption, the energy level splittings of the Tm and Al ions can be drawn as in Figure~\ref{fig:ext}. The two Tm levels split linearly with magnetic field, while the Al levels are constant. Extinction points can occur when the Tm $+1/2\rightarrow -1/2$ transition is resonant with one of the Al transitions. The two extinction points in Figure~\ref{fig:ext001} correspond to the first two Al transitions, the $\pm 1/2 \rightarrow \pm 3/2$ transition at $\Delta_Q$ and the $\pm 3/2 \rightarrow \pm 5/2$  transition at $2\Delta_Q$. A third extinction point is expected at $3\Delta_Q$ when the Tm splitting crosses the third Al transition $\pm 1/2 \rightarrow \pm 5/2$, although this transition is very weak so this point is likely to have less effect on the Tm lifetime than the other two extinction points.

The Al quadrupole splitting $\Delta_Q$ can be written
\begin{equation}
	\Delta_Q = \frac{6Q_p}{4I(2I-1)}
\end{equation}
 where $Q_p$ is the quadrupole parameter. Al occupies two sites in YAG with very different quadrupole splittings: an octahedral ($C_{3i}$ symmetry) site with $Q_p = 0.632$~MHz and a tetrahedral site ($S_4$ symmetry) with $Q_p = 6.017$~MHz \cite{brog66}. The extinction points are caused by cross relaxation with the tetrahedral site, which has $\Delta_Q = 0.90$~MHz. This is in good agreement with the experimental position of the extinction point at a Tm splitting of  $0.91\pm0.02$~MHz, measured from the holeburning spectrum in Figure \ref{fig:ext001}. 

Extinction points like those in Figure~\ref{fig:ext001} can be seen for any direction of the magnetic field. They occur at fixed Tm ground state splittings, equal to the Al splittings. However, because the nuclear Zeeman tensor of Tm in YAG is highly anisotropic, the magnetic field values at which the extinction points occur vary with the field direction.  Since one component of the Tm ground state Zeeman tensor ($\gamma_y \approx 403$~MHz/T) is more than twenty times larger than the other two components, the ground state splitting can be approximated from the projection of this $y$ component onto the magnetic field direction. For the [001] orientation of the magnetic field in Figure~\ref{fig:ext001}, the four sites contributing to the structure each have their $y$ axis 45$^\circ$ from the magnetic field direction. The Zeeman splitting rate of the ground state is therefore $\frac{\Delta_g}{B} = \frac{\gamma_y}{\sqrt{2}}$, or 285 MHz/T. This gives splittings of 0.88 and 1.77~MHz at the two extinction points, in agreement with the observed splittings.

\begin{figure}
	\centering
	\includegraphics[trim = 0 0 8 19,clip]{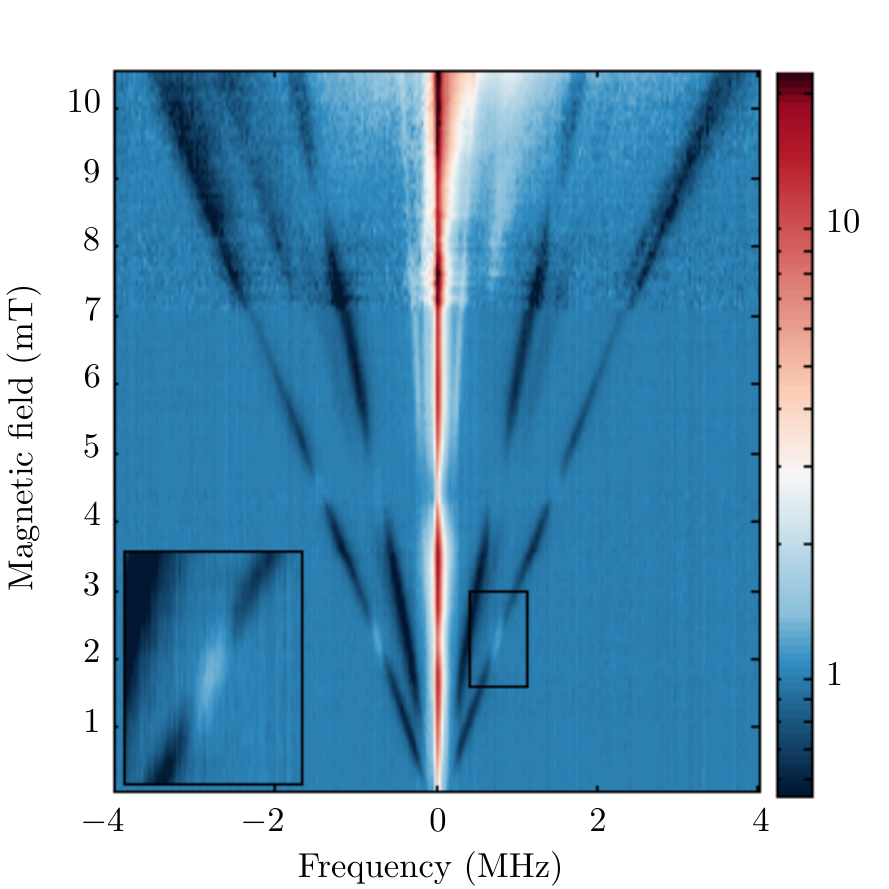}
	\caption{\label{fig:ext110}(Color online) Holeburning transmission spectrum in \tmyag as a function of magnetic field along [110]. The color scale is logarithmic. The inset is a magnified version of the region inside the rectangle: the antihole structure around the first extinction point. A clear inversion of the antihole is seen, which is explained in Section \ref{sec:struct}.  Horizontal slices below 7.1~mT are 8-shot averages as described in the caption of Figure~\ref{fig:ext001}. Extinction of the spectral hole occurs at 2.25~mT and 4.5~mT, 6.75~mT and $\approx 8.8$~mT. Above 7.1~mT, the magnetic field could not be automatically switched off between shots to erase the structure, so each spectrum is a single shot, and the magnetic field was switched off manually between each spectrum. The additional features seen in these spectra are artifacts of this manual method.}
\end{figure} 
Figure~\ref{fig:ext110} shows the holeburning structure for a different magnetic field direction, [110]. In this direction, as for [001], sites 3 to 6 are equivalent.  The laser beam orientation was again chosen to eliminate site 2 but for this case, the polarization was chosen along [110], which allows site 1 to be seen. The spectrum, therefore, displays two sets of sideholes and antiholes. The inner antiholes belong to sites 3 to 6 while the outer antiholes belong to site 1. For this magnetic field direction the strongest antihole, at $\Delta_g-\Delta_e$, is visible for both sites 3 to 6 and site 1 while one of the weaker antiholes, at $
\Delta_g$, is only visible for sites 3 to 6.

The two sets of antiholes display different extinction behavior. The outer, site 1 antiholes disappear at 2.25 and 4.5~mT, while the inner, sites 3 to 6, antiholes disappear at 4.5 and 8.8~mT. Therefore, the extinction point at 4.5~mT is an extinction point for both types of sites, while the 2.25~mT point is only an extinction point for site 1 and the 8.8~mT one only for sites 3 to 6. This behavior can be understood by considering the splitting of these two sets of sites. The $y$ axis of the Zeeman tensor for site 1 is directed along [110], the magnetic field direction, so the ground state splitting rate takes its maximal value $\frac{\Delta_g}{B} = \gamma_y$. Meanwhile, sites 3 to 6 have their $y$ direction 60$^\circ$ from the magnetic field, so for these sites $\frac{\Delta_g}{B} = \frac{1}{2}\gamma_y$. Therefore, at 2.25~mT, site 1 is resonant with the lower Al splitting $\Delta_Q$, while at 4.5~mT, site 1 is now resonant with the higher splitting $2\Delta_Q$ and sites 3 to 6, which have a ground state splitting exactly half that of site 1, are resonant with the lower splitting $\Delta_Q$. Finally, at $\approx 8.8$~mT sites 3 to 6 are resonant with $2\Delta_Q$.

Careful examination of Figure~\ref{fig:ext110} also shows that the amplitude of the site 1 antihole decreases at 6.75~mT, to approximately 30 \% of its original size. At this magnetic field, the Tm ground state splitting is 2.7~MHz, resonant with the weak $3\Delta_Q$,  $\pm\frac{1}{2}\rightarrow\pm\frac{5}{2}$ transition. 

Two things should be noted from the holeburning spectra of Figures~\ref{fig:ext001} and \ref{fig:ext110}. First, there is no visible deviation of the position of the antiholes near the resonance points. This indicates that the interaction is sufficiently weak that it does not substantially perturb the energy level structure of the Tm and Al ions near resonance. Second, the holeburning extinction occurs over a very large range of Tm transition frequencies, of the order of 200~kHz. Since the interaction itself is too weak to cause such a wide extinction range, this demonstrates that other factors, in particular the Tm inhomogeneous broadening of $\mathcal{O}(100)$~kHz, influence the holeburning structure near the resonance.

\section{Measurement of hyperfine lifetimes using stimulated echoes} \label{sec:lifetimes}
In the previous section, we showed that the hyperfine lifetime is reduced at points of resonance between Tm and Al, resulting in a reduction in the central hole and antihole amplitude. The value of the lifetime, at and away from resonance, could be determined from a holeburning spectrum by measuring the hole amplitude as a function of the delay between the holeburning pulse and the readout chirp. However, in a spectral holeburning sequence, the distribution of ions amongst the states of the system at the end of the holeburning pulse is highly dependent on the hyperfine lifetime itself, meaning that ions at resonance and those away from resonance are prepared in completely different ways by the holeburning. This makes analyzing the hole decay somewhat complicated at this stage. An easier method is to use a stimulated (3-pulse) echo sequence, and analyze the echo amplitude as a function of the delay between the second and third pulses. In this sequence the first two pulses prepare the system in a consistent state, greatly simplifying the extraction of the hyperfine lifetime. In this section, we present the theory and experimental data for stimulated echoes at, and away from, the resonance.

\subsection{Relaxation dynamics} \label{sec:lifetimesa}
\begin{figure}
	\begin{center}
	\includegraphics{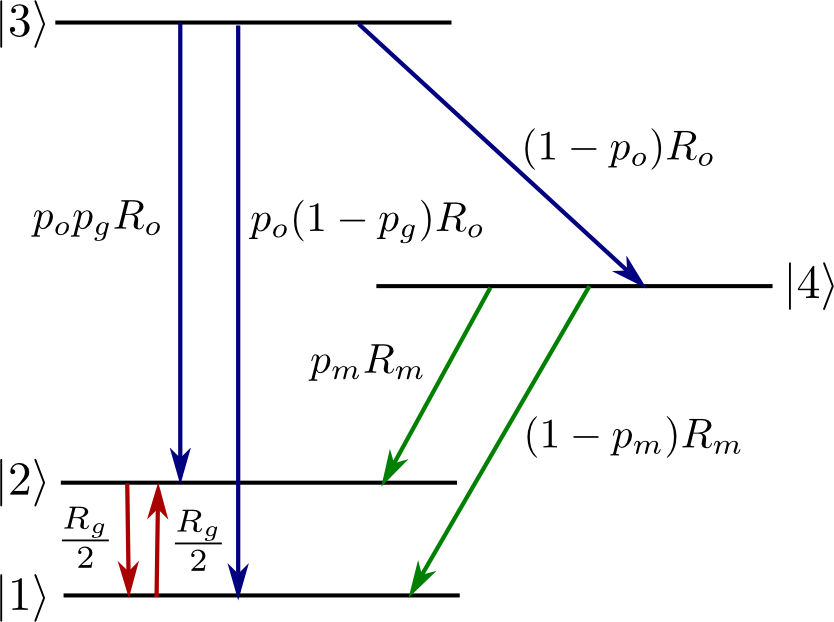}
	\caption{\label{fig:ratepicture} (Color online) 4-level system used in the rate equation model. $R_o$, $R_m$ and $R_g$ are the decay rates of the optical, metastable, and ground states respectively while $p_o$, $p_m$ and $p_g$ are the branching ratios that describe how each of these levels decays. }
	\end{center}
\end{figure}
  To determine the hyperfine lifetime from a stimulated echo decay in a multilevel system like \tmyag, it is necessary to have a model for the shape of the decay curve as a function of the lifetimes and branching ratios of each of the levels involved in the decay process. For \tmyag, a rate equations model with four levels -- two hyperfine ground states, one excited state, and a metastable state -- is sufficient. We are interested in the decay behavior so we consider a set of rate equations with no driving terms. This four level system is illustrated in Figure~\ref{fig:ratepicture}. We will start by assuming that the decay rates are the same for all atoms in the system, which applies when the atoms are far from an extinction point. We will consider the situation when this assumption does not apply later.
  
 The rate equations for the system are:
\begin{align}
	\begin{split}
		\frac{dn_1}{dt}& = \frac{R_g}{2}(n_2-n_1)+p_o(1-p_g)R_on_3+(1-p_m)R_mn_4\\
		\frac{dn_2}{dt}& = -\frac{R_g}{2}(n_2-n_1)+p_op_gR_on_3+p_mR_mn_4\\
		\frac{dn_3}{dt}& = -R_on_3\\
		\frac{dn_4}{dt}& = (1-p_o)R_on_3-R_mn_4
	\end{split}
\label{eq:rateeq}
\end{align}
  Where the $R$ parameters are transition rates and the $p$ parameters branching ratios, as defined in Figure~\ref{fig:ratepicture}. The solution to this linear system of equations is a sum of four exponentials with decay rates of 0, $R_g$, $R_m$ and $R_o$. The explicit solution is given in Appendix \ref{appendixa}.
  
We now apply these rate equations to a stimulated echo experiment. The stimulated echo pulse sequence used in the experiment involves three co-linear pulses, with the first and second pulses separated by time $\tau$ short compared to the coherence time $T_2$, and the second and third pulses separated by a time $T_w$ much longer than $T_2$.  We consider that the pulses are resonant with the transition $|2\rangle\rightarrow|3\rangle$, and the initial population is evenly distributed between the two ground states, $n_1 =n_2 = 0.5$. The first two pulses engrave a spectral population grating between the levels $|2\rangle$ and $|3\rangle$ with a period given by the pulse spacing $\tau$. The populations of the system can be described by a  four element population vector $\vec{\bm{n}}(t,\nu)$. After the engraving step,
\begin{equation}
\vec{\bm{n}}(\tau,\nu) = \vec{\bm{n}}_{eq}+ \xi(\tau,\nu)\vec{\bm{n}}_{gr}
\label{eq:ngrave}
\end{equation}
where $\vec{\bm{n}}_{eq} = \left[\begin{smallmatrix} \frac{1}{2}, &\frac{1}{2},& 0 ,&0\end{smallmatrix}\right]$ describes the equilibrium population distribution, $\vec{\bm{n}}_{gr} = \left[\begin{smallmatrix} 0, &-1,& 1, &0\end{smallmatrix}\right]$ describes the population inversion caused by the engraving pulses, while the frequency dependence of the grating is given by $\xi(\tau,\nu)$. In the simplest case, this is of the form
\begin{equation}
\xi(\tau,\nu) = \frac{1}{2}\xi_0\left(1+\cos\left(2\pi\nu\tau+\phi\right)\right)
\label{eq:xi}
\end{equation}
for frequencies within the grating bandwidth. $\xi_0\ll1$ is the grating depth.

During the wait time between the second and third pulses, the grating will degrade. The evolution of $\vec{\bm{n}}(t,\nu)$ after the engraving step can be written as
\begin{equation}
\vec{\bm{n}}(t,\nu) = \bm{U}(t-\tau)\vec{\bm{n}}(\tau,\nu)
\label{eq:nev}
\end{equation}
where the evolution matrix $\bm{U}$ can be derived from the solution to the rate equations (Equations \eqref{eq:rateeq}). Since the evolution matrix does not affect the equilibrium population $\vec{\bm{n}}_{eq}$, the population can then be written
\begin{equation}
\vec{\bm{n}}(t,\nu) = \vec{\bm{n}}_{eq}+ \xi(\tau,\nu) \bm{U}(t-\tau)\vec{\bm{n}}_{gr}
\end{equation}
Thus, the evolution of the grating is described simply by the evolution of $\vec{\bm{n}}_{gr}$ under $\bm{U}$.

The third pulse of the stimulated echo sequence scatters off the spectral grating, forming an echo at time $\tau$ after the third pulse. The echo amplitude is proportional to the visibility of the spectral grating at the time of the third pulse, and can be written
\begin{equation}
A(T_w) \propto \vec{\bm{n}}_{gr}^\intercal \bm{U}(T_w) \vec{\bm{n}}_{gr}
\end{equation}
The echo intensity, which is the parameter measured in the experiment, is then
\begin{equation}
I(T_w) = C_0 \left(\vec{\bm{n}}_{gr}^\intercal \bm{U}(T_w) \vec{\bm{n}}_{gr}\right)^2
\label{eq:dec}
\end{equation}
where $C_0$ is a proportionality constant. Since $\bm{U}$ is derived from the rate equations, there are seven unknowns in this equation: $C_0$, the three decay rates $R_o$, $R_m$, and $R_g$, and the three branching ratios $p_o$, $p_m$ and $p_g$. Of these, $p_g$, which is the branching ratio from the excited state to the initial hyperfine ground state, can be fixed at $p_g = 1$. In general, $p_g$ is the same as the branching ratio of excitation from the hyperfine ground states to the excited state if the decay is direct or if the time spent in any intermediate levels is short, which is the situation that applies to \tmyag. Due to the highly anisotropic Zeeman interaction in \tmyag, the excitation branching ratio is close to one along all but a few special directions.

\subsection{Decay away from resonance} \label{sec:lifetimesb}
Equation \eqref{eq:dec} gives a model of the echo decay curve as a function of six parameters. To extract these parameters we measured the stimulated echo decay at a magnetic field away from any extinction points. All measurements were performed in a helium bath cryostat with the sample temperature maintained below 2.6~K. Close control of the temperature is necessary for \tmyag as the hyperfine ground state lifetime is temperature dependent above approximately 2.8~K. An external coil was used to supply a magnetic field of 4.56~mT parallel to the [001] direction. The laser beam was oriented along $[\bar{1}10]$ and polarized along [001], which, as noted in the previous section, means that only the four equivalent sites 3 to 6 interact with the laser.

The laser power was 1.6~mW, focused onto the crystal. In the first part of the stimulated echo sequence, two 1~\textmu s square pulses separated by 5~\textmu s were used to create the population grating, which was read out with a third, identical pulse at delays between 100~\textmu s and 60~ms. The repetition rate was slow, 0.3~Hz, and after every shot the magnetic field was switched to the extinction point at 3.1~mT for 200~ms to erase the population grating.

\begin{figure}
	\begin{center}
	\includegraphics[trim = 0 0 10 10,clip]{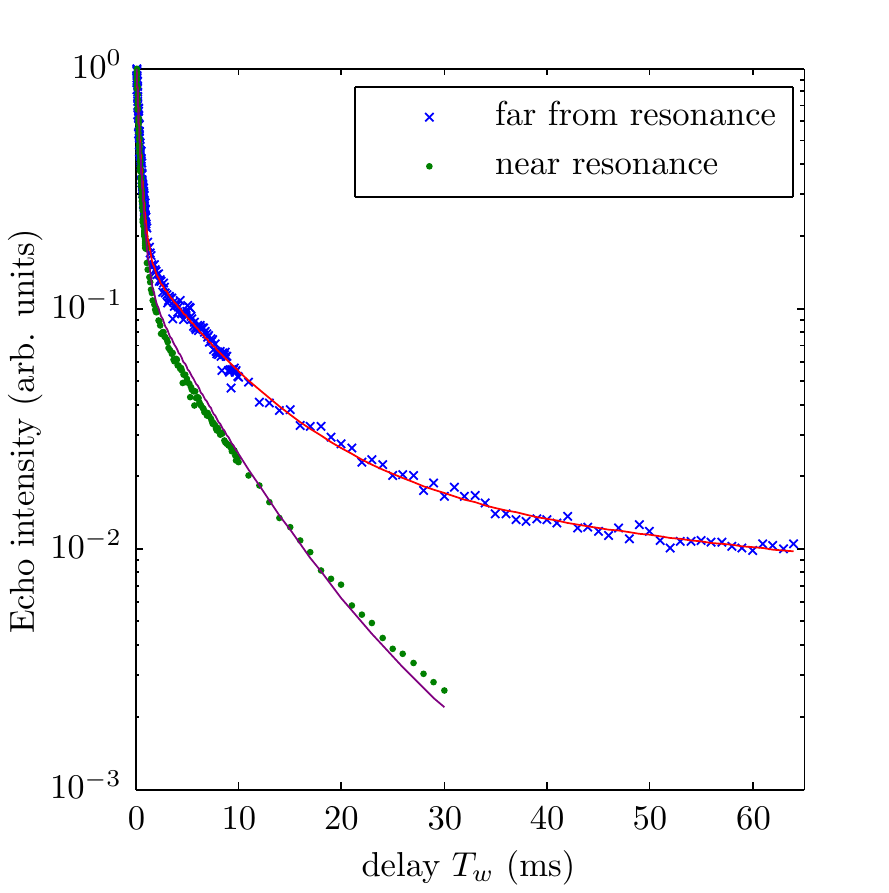}
	\caption{\label{fig:echo65} (Color online) Stimulated echo intensity decay for magnetic fields along the [001] direction far from resonance (4.56~mT, blue crosses), and close to resonance (6.2~mT, green dots). $T_w$ is the delay between the second and third pulses of the sequence. The red line is a fit to the far from resonance data using Equation \eqref{eq:dec}, while the purple line is a fit to the near resonance data using Equation \eqref{eq:exmodel}.}
	\end{center}
\end{figure}
The decay of the stimulated echo intensity is shown in Figure~\ref{fig:echo65}. Also shown is a fit to the data using the model of Equation \eqref{eq:dec}. Three decay regions can be identified in the figure. At  first, $T_w<2$~ms, the curve is dominated by the rapid excited state decay, with a rate $R_o = (2.05\pm0.04) \times 10^3$~s$^{-1}$ ($T_1^o = 0.49\pm0.01$~ms). For intermediate $2<T_w<30$~ms, the metastable state decay, with $R_m =94\pm4$~s$^{-1}$ ($T_1^m = 10.6\pm0.4$~ms) dominates, while at larger times, the slow ground state decay, with $R_g = 4.4\pm0.8$~s$^{-1}$ ($T_1^g = 230\pm40$~ms) begins to become important. The fit to the data also gives the branching ratios $p_o = 0.30\pm0.02$ and $p_m = 0.67\pm0.01$. As stated above, $p_g$ is fixed at 1. 

The fitted values for the lifetime of the optical and metastable levels, 0.49~ms and 10.6~ms respectively, are in good agreement with previously published values \cite{caird75, macfarlane93, strickland00}.The hyperfine level lifetime is considerably shorter than the lifetimes of several seconds that can be seen in holeburning measurements at similar fields. This suggests that some degree of spectral diffusion is contributing to the decay process. We chose not to add spectral diffusion to our model, because the only parameter affected by this inclusion is the hyperfine lifetime, and it was not a priority of our experiment to measure an accurate hyperfine lifetime in the non-resonant regime.

\subsection{Decay near resonance}\label{sec:extlifetimes}

The fitted parameters gained in the previous section for the decay away from resonance can  be used to build a model for the decay at resonance. The crucial difference between these two situations is that at resonance, the hyperfine state decay can no longer be described by a single decay rate. This occurs because the Tm hyperfine transition is inhomogeneously broadened, so that the lifetime of each ion depends on how far away it is from resonance with the Al transition. The ensemble must be described by a distribution of hyperfine lifetimes. The Al transition is also inhomogeneously broadened, but because the Al magnetic moment is small, this broadening is much smaller than the Tm broadening and will be ignored.

To model the effect of the Tm inhomogeneous broadening on the echo near resonance, we first write the hyperfine decay rate $R_g$ as a function of the  distance from resonance with the Al ion:
\begin{equation}
	R_g(\nu) = r_0+(r_{res}-r_0)h(\nu-\nu_{Al})
	\label{eq:rgf}
\end{equation}
where $r_0$ is the decay rate far from resonance measured in the previous section, $r_{res}$ is the rate exactly on resonance, and $h(\nu-\nu_{Al})$ is the lineshape of the resonant interaction between Al and Tm, centered on the Al splitting $\nu_{Al}$. For the moment, we will assume this function is a single Lorentzian with an amplitude of one. In Section \ref{sec:zeeman} we will examine $h$ in more detail. The width of $h$ is $\Gamma_{ff}$, the flip-flop resonance width, which is dependent on the strength of the Al-Tm interaction.

 As the hyperfine level decay rate is now a function of the Tm ground state splitting frequency, the contribution to the echo amplitude for the subset of ions with hyperfine frequency $\nu$ is
\begin{equation}
	A(T,\nu,\Delta_g) = C_0^{1/2}\left(\vec{\bm{n}}_{gr}^\intercal \bm{U}(T) \vec{\bm{n}}_{gr}\right)G(\nu-\Delta_g)
\end{equation}
where $G(\nu-\Delta_g)$ is the inhomogeneously broadened lineshape of the hyperfine transition, centered on the frequency $\Delta_g(B)$. The echo intensity, then, is the square of the integral of $A(T,\nu,\Delta_g)$ over the lineshape $G(\nu-\Delta_g)$:
\begin{equation}
	I(T,\Delta_g) = C_0\left(\int_{-\infty}^{\infty}\left(\vec{\bm{n}}_{gr}^\intercal \bm{U}(T) \vec{\bm{n}}_{gr}\right)G(\nu-\Delta_g)d\nu\right)^2
	\label{eq:exmodel}
\end{equation}
The maximum effect on the echo is seen when the Tm and Al transitions are in resonance, $\Delta_g = \nu_{Al}$. At this point, the overlap between the flip-flop resonance lineshape $h(\nu-\nu_{Al})$ and the inhomogeneous lineshape $G(\nu-\Delta_g)$ is greatest.

Equation \eqref{eq:exmodel} is dependent on the parameters determined in the previous section, as well as three new parameters: the maximum hyperfine decay rate $r_{res}$, the flip-flop resonance width $\Gamma_{ff}$ and the hyperfine inhomogeneous broadening, $\Gamma_{inh}$.  This last parameter can be determined from the holeburning spectrum in Figure~\ref{fig:ext001}. The lineshape of the antiholes in this spectrum is a convolution of the central hole lineshape with the hyperfine inhomogeneous broadening function $G$. A fit to the antihole lineshape gives $G$ as a Gaussian function with $\Gamma_{inh}$ of 120~kHz (FWHM).

The other two parameters, $r_{res}$ and $\Gamma_{ff}$, cannot  be determined solely from a fit to the echo decay curve at resonance because they both have a similar effect on the curve. However, the combination of the echo decay fit and the  fit to the antihole structure near resonance (Section \ref{sec:anticomp}) determine that $r_{res} = (2 \pm 0.8)\times 10^3$~s$^{-1}$ and $\Gamma_{ff} = 12\pm 3$~kHz.

Figure~\ref{fig:echo65} shows the echo decay for the 1.8~MHz extinction point at 6.2~mT along the [001] direction. The echo decays much more rapidly than the far-from-resonance curve shown in the same figure, and has only two clear decay regions: at short times, the decay due to the excited state and at longer times, that due to the metastable state. The purple line in the graph shows the model prediction of Equation \ref{eq:exmodel},  using $\Gamma_{inh} = 120$~kHz, $r_{res} = 2\times 10 ^3$~s$^{-1}$ ($T_1^g(res) = 0.5$~ms), $\Gamma_{ff} = 12$~kHz, and the decay rates and branching ratios from the far-from-resonance fit.  This calculated $T_1$ for the hyperfine level is substantially shorter than the metastable state lifetime, and of the order of the excited state lifetime.

In Figure~\ref{fig:echo65}, the model of Equation \ref{eq:exmodel} matches the experimental decay curve well. There is a slight discrepancy at medium to long times, which can be attributed to the extreme simplicity of the model used. In particular, it was assumed that the effect on the hyperfine lifetime of the resonance with the Al could be represented by a single Lorentzian fixed at the center of the resonance. This assumption only holds if the Al Zeeman structure is ignored, and we show in Section \ref{sec:zeeman} that accurately modelling the behavior near resonance requires taking the Zeeman structure into account.

\section{Modelling the holeburning structure} \label{sec:micromodel}
In the previous section, we showed that the stimulated echo decay near resonance could be well explained by a model that took into account the inhomogeneous broadening of the Tm hyperfine transition. One of the implicit assumptions in this model was that the inhomogeneous broadening is much greater than the Tm-Al flip flop resonance width. In the present section, we substantiate this assumption by investigating the holeburning structure near resonance. 

As previously mentioned, the holeburning structure is slightly difficult to model because both the preparation and readout steps of the experiment are highly dependent on the hyperfine lifetime, unlike the echo decay, for which only the readout step is lifetime dependent. To accurately model the holeburning structure, therefore, we make use of the lifetimes and the branching ratios calculated in the previous section from the stimulated echo decays. We will concentrate on the anti-hole structure, as, for reasons we will shortly explain, this is more sensitive to the details of the resonant Tm-Al interaction than either the hole structure or the echo decay.

\subsection{Antihole and hole structure} \label{sec:struct}
To model the hole and antihole structure, we use the four level system of Figure~\ref{fig:ratepicture}, where the driving field is again tuned to the $|2\rangle \rightarrow |3\rangle$ transition.  We assume that the holeburning step is not saturated, so that the shape of the spectral hole $\zeta(\nu-\nu_0)$ centered around the optical frequency $\nu_0$ is invariant with the distance from resonance $\Delta_g-\nu_{Al}$. Then, only the amplitude of the spectral hole changes with $\Delta_g-\nu_{Al}$. We begin with the structure of the central hole and the anti-hole for ions with a fixed ground state splitting $\Delta_g$.

The central hole amplitude for a particular ground state splitting is dependent on the population $n_2$ at the end of the holeburning and wait steps. The population after the holeburning step $n_2(t_b,\Delta_g-\nu_{Al})$ can be calculated from a rate equations model similar to Equations \eqref{eq:rateeq} with a driving term added, in which the ground state relaxation rate is allowed to vary as in Equation \eqref{eq:rgf}. Then, the population after the wait step $n_2(t_w,\Delta_g-\nu_{Al})$ can be determined by applying the solution to Equations \eqref{eq:rateeq} with the same varying $R_g$ to $n_2(t_b,\Delta_g-\nu_{Al})$. The logarithm of the hole amplitude is then proportional to
\begin{equation}
	k_h(\Delta_g-\nu_{Al}) = n_{eq}-n_2(t_w,\Delta_g-\nu_{Al})	
	\label{eq:kh}
\end{equation}
where $n_{eq} = 0.5$ is the equilibrium population of the level. A similar expression exists for the antihole:
\begin{equation}
	k_{ah}(\Delta_g-\nu_{Al}) = n_{eq}-n_1(t_w,\Delta_g-\nu_{Al})	
	\label{eq:kah}
\end{equation}

Equation \ref{eq:kh} gives the contribution of ions at a particular frequency $\Delta_g$ to the hole amplitude. If we now take into account the inhomogeneous broadening of the hyperfine ground state, the hole structure  for an average ground state splitting of $\Delta_g$ is given by integrating $k_h$ over the inhomogeneous lineshape 
\begin{equation}
H_h(\nu,\Delta_g) =  \zeta(\nu-\nu_0)\int_{-\infty}^{+\infty}  k_h(\Delta_g+\alpha-\nu_{Al})G(\alpha)d\alpha
\label{eq:holeamp}
\end{equation}
The integral is simply a convolution of $k_h$ and $G$. This convolution means that the hole amplitude is reduced over a range of Tm hyperfine splittings comparable to the inhomogeneous broadening,  a range much bigger than the width of $k_h$ itself. This explains why the experimental holeburning spectra in Figures \ref{fig:ext001} and \ref{fig:ext110} show holeburning extinction over such large regions. A second important consequence of the convolution is that any structure in $k_h$ smaller than the inhomogeneous broadening will not be visible in $H_h$.

In contrast to the central hole, the anti-holes are sensitive to the structure of the resonance. To prove this, we consider the antihole at $\Delta_g$ as the equations are somewhat simpler for this case. We present results for the antihole at $\Delta_g-\Delta_e$, the one most commonly seen in the experiment, in the Appendix \ref{appendixb}. 

The antihole centered at an optical frequency of $\nu_0-\Delta_g$ can be described by
\begin{equation}
H_{ah}(\nu,\Delta_g) =- \int\limits_{-\infty}^{+\infty} k_{ah}(\delta\nu+\alpha-\nu_{Al})G(\delta\nu+\alpha-\Delta_g)\zeta(\alpha)d\alpha
\label{eq:aholeamp}
\end{equation}
where $\delta\nu = \nu-\nu_0$ is the optical detuning. In this case, the convolution is between the hole lineshape and the \emph{product} of $k_{ah}(\delta\nu-\nu_{Al})$ and $G(\delta\nu -\Delta_g)$.  Because the functions $k_{ah}$ and $G$ are multiplied and not convoluted as for the structure of the hole (Equation \eqref{eq:holeamp}), the antihole near the resonance will be sensitive to any structure in $k_{ah}$ larger than the width of $\zeta$. 

Equation \eqref{eq:aholeamp} predicts that in a two-dimensional holeburning spectrum such as Figure \ref{fig:ext110} the resonance point will appear as a vertical line of reduced antihole amplitude which is  separated from the central hole by the Al splitting and whose  vertical extent is given by the inhomogeneous broadening. This broadly agrees with the appearance of antiholes in the figure.  

Many of the extinction points in Figure \ref{fig:ext110} show not just reduction but inversion of the antihole. This  is most clearly seen in the inset, a magnified view of the first extinction point at 2.25~mT. The inversion can be explained by considering the dynamics of the holeburning process. At resonance, we have shown that the hyperfine state lifetime is shorter than the metastable state lifetime. Importantly, it is also shorter than the holeburning pulse length, meaning that the two hyperfine levels can come into equilibrium during the holeburning pulse. As the resonant level is continually being emptied by the laser, the relaxation between the two hyperfine levels will begin to empty the non-resonant one.  Therefore, at resonance the holeburning process pumps population out of both hyperfine levels to be stored in the metastable state, burning a sidehole and not an antihole in the non-resonant level.

Another aspect of the antihole structure that is visible in the inset of Figure \ref{fig:ext110}  is that the inverted region is diagonal, and not vertical. This diagonal shape indicates that the antihole extinction function $k_{ah}$ is structured, made up of multiple resonances whose separations are larger than the individual resonance width. The diagonal inverted structure, therefore, is made up of a number of offset vertical resonances as described by the Equation \ref{eq:aholeamp}. Multiple closely spaced resonances arise in \tmyag when we take into account the Zeeman structure of Al, which is the topic of the next section.

\subsection{Role of the Al Zeeman splitting} \label{sec:zeeman}
Thus far we have ignored the effect of the small Al Zeeman splitting in order to simplify the models for the holeburning structure and stimulated echo decay. We saw in the previous section that the Al Zeeman splitting does have an effect on the antihole structure near resonance, so in this section we include it into the model for the antihole structure.

Ignoring the Al Zeeman interaction substantially simplified the models of the previous sections because it is equivalent to assuming that all Al ions in the crystal have the same hyperfine structure. When we take the Al Zeeman tensor into account, this is no longer the case. While the Al Zeeman tensor, which has $\gamma_{Al} = 11.1$~MHz/T, is isotropic, when combined with the anisotropic quadrupole tensor it leads to energy levels splittings that are dependent on the orientation of the quadrupole axis with respect to the applied magnetic field.  This means that Al ions with different orientations will have different energy level structure, and therefore lead to differently structured extinction points when they interact with Tm.

 \begin{figure}
	\begin{center}
	\includegraphics[trim = 0 0 8 10,clip]{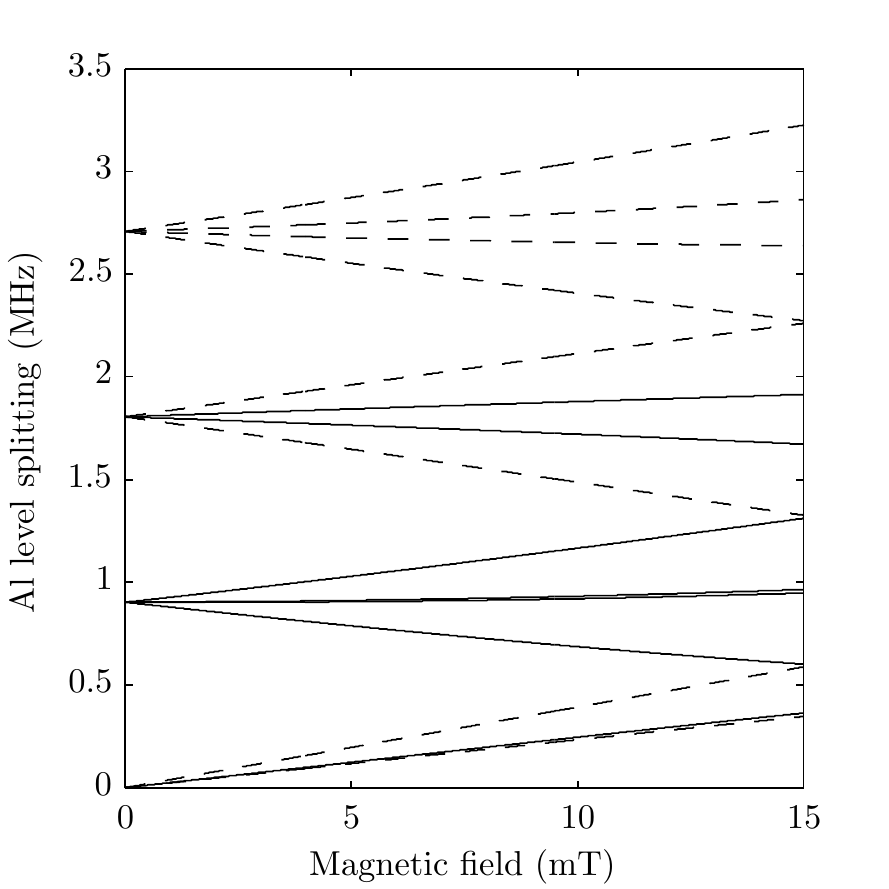}
	\caption{\label{fig:split45} Hyperfine transition frequencies of Al in YAG for a magnetic field  $45^\circ$ from the direction of the quadrupole quantization axis. Solid lines indicate strong transitions (transition probability above 0.1) while dashed lines indicate weak transitions (probability below 0.01).}
	\end{center}
\end{figure}
The  tetrahedral Al ions in YAG have three different orientations  for the quadrupole axis, along the three (001) directions \cite{alff61}. For magnetic fields along [001] and [110], there are, therefore, three possible orientations of the axis relative to the field: parallel, perpendicular, and at an angle of $45^\circ$. Each will lead to a different pattern of splitting of the Al hyperfine levels. As an example, we show the splitting for the case of $45^\circ$ in Figure~\ref{fig:split45}. The magnetic field lifts the double degeneracy of all three quadrupole levels. A similar pattern with larger splitting is seen for a field parallel to the quadrupole axis, while for a field perpendicular to the axis only the $m_z = \pm \frac{1}{2}$ levels split.

Since the Al splittings are now dependent on the orientation of the Al sites, the extinction point structure will be dependent on which Al ions a particular Tm site interacts with. This can be determined from the crystal structure. Tm in YAG occupies six magnetically inequivalent sites of D$_2$ symmetry (3 mutually perpendicular $C_2$ symmetry axes). The difference between these six sites is the orientation of their symmetry axes. Each pair of sites (1 and 2, 3 and 4, 5 and 6) has one $C_2$ axis along a (001) direction, and the other two along perpendicular (110) directions. We will denote the (001) direction of a site as the polar axis.

 Each Tm ion has two nearest neighbor tetrahedral Al ions at 3.0~\AA\ in either direction along the polar axis \cite{euler65,shelyapina06}. The quadrupole axis for these two Al ions is also along the polar axis. Thus,  for site 1 and 2 Tm ions the nearest neighbor Al ions have their quadrupole axis along [001], for sites 3 and 4 along [100] and for sites 5 and 6 along [010].

 The next nearest neighbor  tetrahedral Al ions are 4 ions at  a distance of 3.6~\AA. These 4 ions divide into two pairs with quadrupole axes along the two (001) directions perpendicular to the polar axis.  The third nearest neighbors are considerably further away than the next nearest neighbors, at 5.6~\AA, and will not be considered here.

 \begin{figure}
	\begin{center}
	\includegraphics[trim = 0 0 8 10,clip]{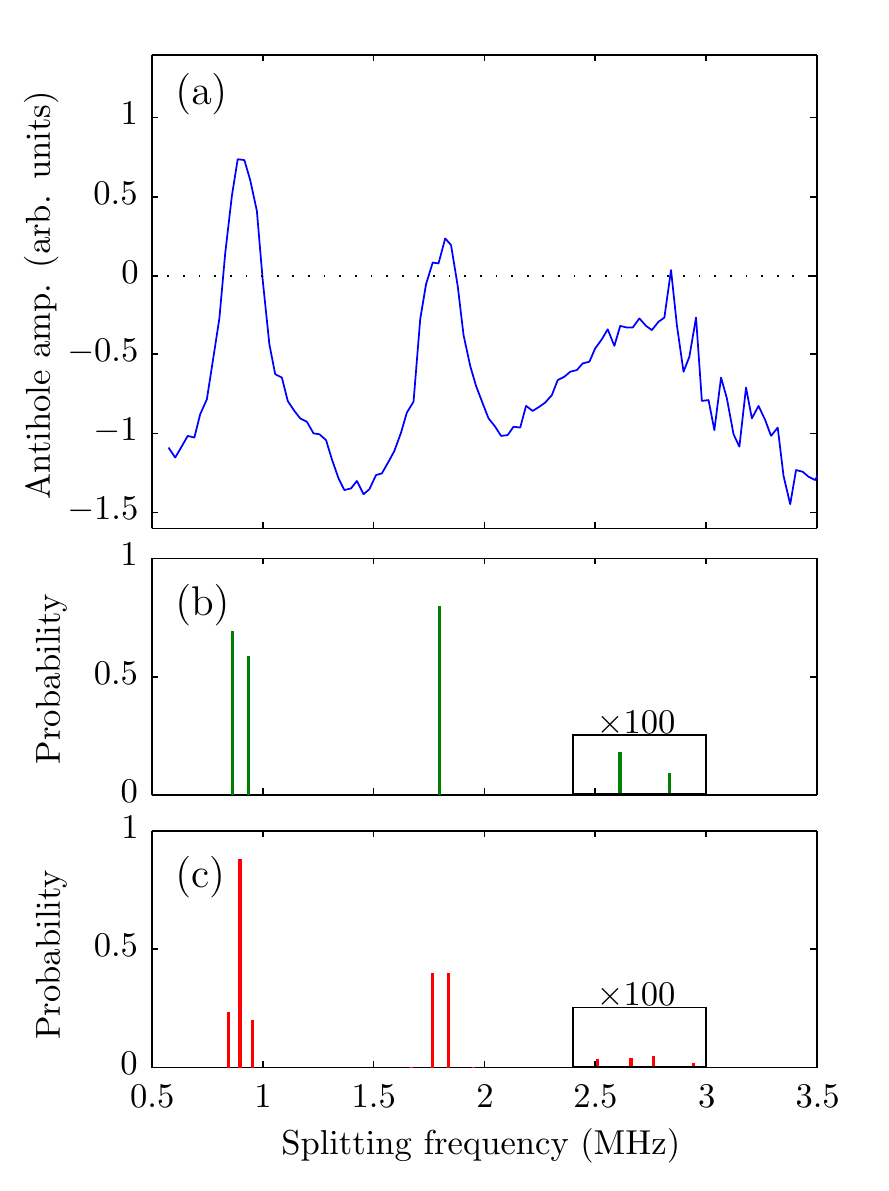}
	\caption{\label{fig:comp1101} (Color online) (a) Site 1 antihole amplitude as a function of the Tm ground state splitting for a magnetic field along [110]. All three Al resonances, at 0.9, 1.8 and 2.7~MHz, are visible. (b) Position of the resonant Al transitions for the two nearest neighbor ions to site 1, whose polar axis is perpendicular to the field. The probabilities for the transitions above 2.5~MHz are multiplied by 100 to make them visible.   (c) Position of resonant Al transitions for the four next nearest neighbor ions ( quadrupole axis $45^\circ$ to the field).  }
	\end{center}
\end{figure}
 To show how this information can be used to determine the Al resonance frequencies for a particular Tm site, we shall use the extinction point at 2.25~mT in Figure \ref{fig:ext110} as an example. In this figure, the magnetic field is along [110] and the extinction point is due to ions in site 1, whose polar axis is [001]. Thus, their nearest neighbour Al ions will have their quadrupole axis along [001] as well, perpendicular to the field, while the next nearest neighbours have their quadrupole axis at 45$^\circ$ to the field. 
 
 Figure \ref{fig:comp1101} compares the energy level splitting for these sets of Al ions (bottom two plots) to the depth of the antihole as a function of magnetic field, top, which was obtained by cutting diagonally across Figure \ref{fig:ext110} through the center of the antihole. In this figure, vertical lines indicate calculated Al hyperfine transition frequencies for the nearest and next nearest neighbour sites. Their amplitudes are given by the corresponding transition probability, which can be calculated for an RF transition from the expectation value of the magnetic dipole operator.
 
 The positions and widths of the extinction points in Figure \ref{fig:comp1101} correlate well with the Al structure: the first two extinction points are relatively narrow due to the small Al splittings, while the third, weaker extinction point is much broader due to the large spread in Al structure. The degree of inversion of the antihole, which decreases with increasing splitting, also agrees with the general decline in Al transition probabilities with increasing splitting.

 \subsection{Comparison between model and  antihole structure}\label{sec:anticomp}
 The calculation of the Al Zeeman structure in the previous section can be used to more accurately model the Tm antihole structure near resonance. To modify the antihole structure model of Section \ref{sec:micromodel} to include the Al structure, it is sufficient to feed an altered function for $R_g$ into the rate equation models used:
\begin{equation}
R_g(\nu) = r_0+(r_{res}-r_0)\left[\sum_i h(\nu-\nu_{Al,i})\right]
\label{eq:rgsum}
\end{equation}
where $\nu_{Al,i}$ are the different Al transition frequencies that contribute to the resonance. 

We will again apply this to the site 1 extinction point for a 2.25~mT magnetic field along [110]. This is the simplest case to model as only one site contributes to the antihole. Since the visible antihole is at $\Delta_g-\Delta_e$, we use the antihole structure equation suitable for this antihole (Equation \ref{eq:hah}). We choose to only consider the effect of nearest neighbour Al sites as these should dominate the interaction. The nearest neighbor ions lead to two resonances $\nu_{Al}$ at 0.86 and 0.94~MHz (see Figure~\ref{fig:comp1101}). The lineshape of the hyperfine decay rate $R_g(\nu)$ is then given by Equation \eqref{eq:rgsum}.

The model is dependent on the  inhomogeneous broadening lineshape $G(\delta\nu-(\Delta_g-\Delta_e))$ and the hole lineshape $\zeta(\nu)$, which can be measured from the holeburning spectrum away from resonance. As previously mentioned, $G$ is a Gaussian function with FWHM 120~kHz, while  $\zeta$ is a Lorentzian with a linewidth of 50~kHz.

\begin{figure}
	\begin{center}
	\includegraphics[trim = 0 0 8 19,clip]{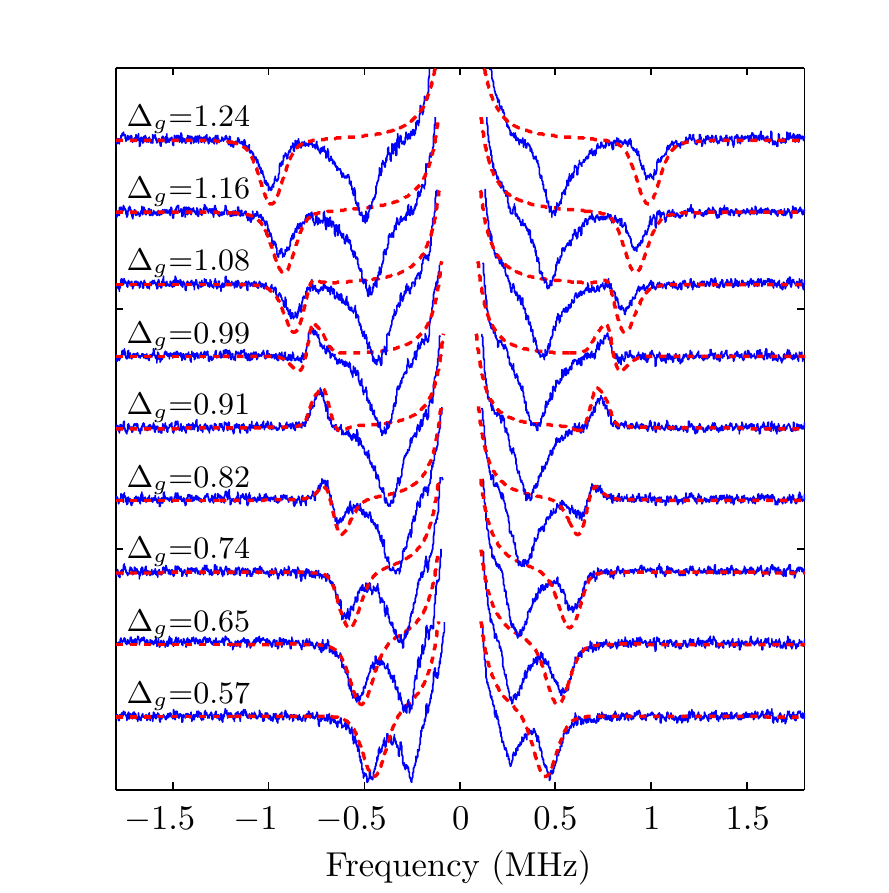}
	\caption{\label{fig:ahmodel} (Color online) Comparison between experimental antihole structure (blue solid lines) and modeled structure (red dashed lines) near the site 1 resonance at 2.25~mT along [110]. The different traces are different Tm ground state splittings, with the value of the splitting listed for each trace in MHz. The resonance is at approximately 0.91~MHz.}
	\end{center}
\end{figure}
The antihole structure is also dependent on $r_{res}$ and $\Gamma_{ff}$, parameters that occur in the model for the stimulated echo decay shape of Section \ref{sec:extlifetimes}. As described in that section, the values of $r_{res}$ and $\Gamma_{ff}$ that best fit both the echo decay and the antihole structure near resonance are  $r_{res} = (2 \pm 0.8)\times 10^3$~s$^{-1}$ and $\Gamma_{ff} = 12\pm 3$~kHz. 

Using the parameter values described above, Figure~\ref{fig:ahmodel} compares the model spectrum to the experimental data for several different values of the ground state splitting near the resonance. The model matches the general behavior well, showing first a partial inversion of the high frequency components of the antihole as these ions come into resonance (0.82~MHz), then a centralized, complete inversion (0.91~MHz), and finally a partial inversion on the low frequency side (0.99~MHz). The amplitude of the antihole far from resonance (0.57~MHz), and the amplitude of the inverted peak are well matched by the model, although the experimental spectrum takes longer to return to the original antihole size, suggesting that the wings of the resonance are slightly wider than the model predicts. This could be due to a contribution from the next-nearest neighbors, which were ignored in the model.  
 
 \begin{figure}
	\begin{center}
	\includegraphics[trim = 0 0 8 10,clip]{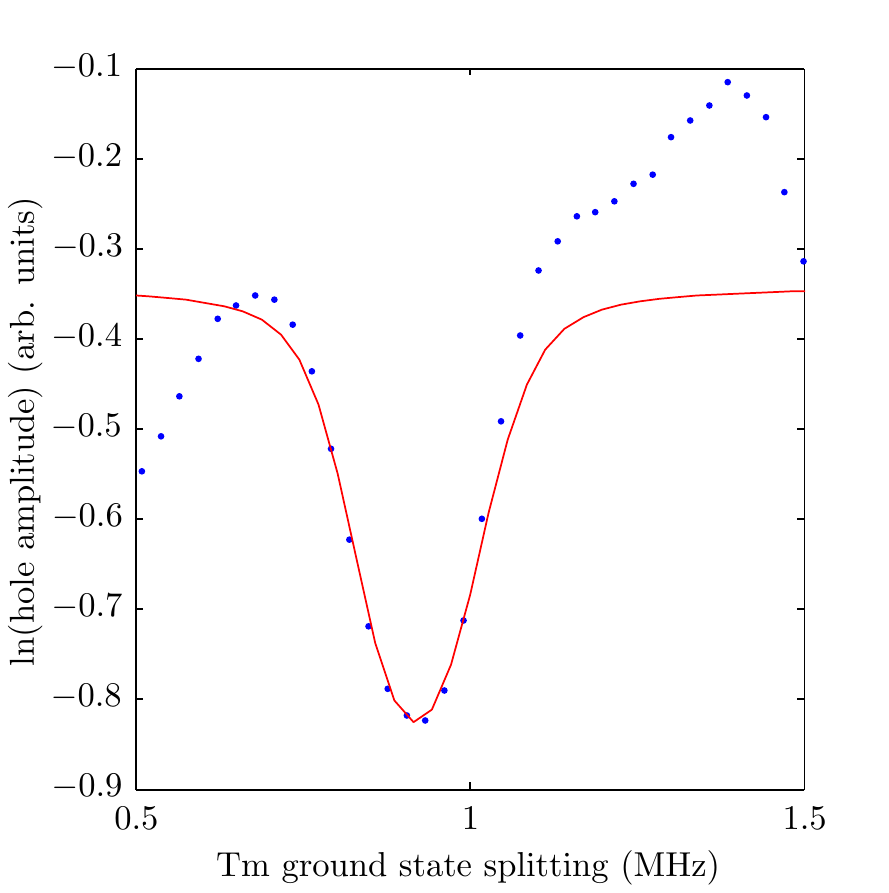}
	\caption{\label{fig:hmodel} (Color online) Central hole amplitude as a function of the Tm ground state splitting for the 2.25~mT extinction point of site 1 along [110]. The experimental spectrum is given by the blue dots while the model of Equation \eqref{eq:holeamp} is shown by the red line. }
	\end{center}
\end{figure}
The same set of parameters as used above can be used to model the central hole structure near the resonance, based on Equation \eqref{eq:holeamp}. Figure~\ref{fig:hmodel} shows the  experimental amplitude of the central hole compared to that of the model as a function of the Tm ground state splitting. The resonance is relatively shallow because half the signal comes from ions in sites 3 to 6, which are not at an extinction point. The model is a close fit to the experimental data in the region of the resonance. Outside the resonance region, the experimental amplitude increases with the splitting, an indication that the lifetime is lengthening. This shows that there are other factors affecting the lifetime apart from the Al-Tm cross-relaxation resonances. 

\section{Discussion}\label{sec:discussion}

We have shown that, when the Tm and Al nuclear splittings coincide, relaxation of the Tm spin happens rapidly through a resonant flip-flop interaction with nearby Al ions. The relaxation is completed within a few milliseconds, meaning that Al thermalization rate must be consistent with that time scale. 


The Tm-Al flip-flop is expected to be driven by the nuclear magnetic dipole-dipole interaction. This interaction can be modeled using the well-known crystal structure of YAG and the Tm and Al gyromagnetic tensors. Understanding the relaxation process requires, in addition, some model for the stochastic flipping of the Al spins, which results from their coupling to each other. The interactions between Al ions are likely to be affected by the existence of a small frozen core around the Tm ion arising from the enhanced nuclear magnetic moment of Tm. This frozen core complicates the model of the Al spin bath dynamics, as it means that Al near the Tm relax at different rates to those further away. We defer the theoretical analysis of these spin bath dynamics, and the resulting Tm-Al flip-flop rate, to a future paper.

For the practical applications of \tmyag to quantum and classical information processing, the resonant Tm-Al interactions studied here could prove useful for manipulating the Tm hyperfine level populations. Switching the magnetic field to an extinction point provides a way of equalising the populations of the two levels without reducing the field completely to zero. It is also possible to manipulate the hyperfine states of the different Tm sites independently, as they have extinction points at different fields.

\section{Conclusion}
At well defined magnetic fields $<10$~mT, the lifetime of the \tmyag hyperfine ground states decreases substantially, which is seen as a reduction in the holeburning amplitude. This was attributed to a resonant cross-relaxation interaction between Tm and Al. We studied the relaxation dynamics both at and away from the resonance using stimulated echoes.

Away from resonance at a magnetic field of 4.56~mT, the echo decay was well described with a single hyperfine decay rate with $T_1 = 230$~ms. At a resonance, the inhomogeneous broadening of the Tm hyperfine transition means that the ions have a range of different distances from resonance and therefore different decay rates. We showed that a model taking into account these different rates is a good fit to the observed echo decay curve for a minimum hyperfine lifetime of $0.5\pm0.3$~ms and a cross-relaxation interaction width of $12\pm3$~kHz.

We also studied the spectral holeburning structure near resonance. For most of the resonances between Al and Tm, the antiholes become inverted at resonance, a clear sign that the hyperfine lifetime is much smaller than the metastable state lifetime. We modeled the structure of the antiholes and the central hole near resonance with a model that included the effect of the hyperfine inhomogeneous broadening as well as the Zeeman structure of Al. In a future theoretical work, we shall model the cross-relaxation process, considering both the transfer of excitation to the neighboring Al ions and the dissipation of that excitation to the environment, to see if the observed behaviour is consistent with the dipole-dipole interaction commonly invoked for cross-relaxation interactions between nuclear spins.

This experimental investigation has dealt with the spontaneous flip flop between Tm and Al. A controlled flip flop can be achieved using  the cross relaxation technique proposed by Hartmann and Hahn \cite{hartmann62}. This kind of approach, combined with optical detection, could be applied to the present system.

\begin{acknowledgements}
The research leading to these results has received funding from the People Programme (Marie Curie Actions) of the European Union's Seventh Framework Programme FP7/2007-2013/ under REA grant agreement no. 287252 and from the national grants ANR-12-BS08-0015-02 (RAMACO) and ANR-14-CE26-0037-02 (DISCRYS)
\end{acknowledgements}

\appendix
\section{Solution to the 4-level rate equations}\label{appendixa}
The 4-level rate equations in Equations \eqref{eq:rateeq} are straightforward to solve. The solution for the population vector $\vec{\bm{n}} =  \left[ n_1, n_2, n_3 ,n_4\right]$ 
\begin{equation}
	\vec{\bm{n}} = c_1\vec{\bm{x}}_1+c_2\vec{\bm{x}}_2e^{-R_gt} + c_3\vec{\bm{x}}_3e^{-R_mt}+c_4\vec{\bm{x}}_4e^{-R_ot}
\end{equation}
The eigenvectors $\vec{\bm{x}}$ are:
\begin{equation}
	\vec{\bm{x}}_1 = \left[\begin{array}{c} 
	1\\1\\0\\0\end{array}\right]
\end{equation}
\begin{equation}
	\vec{\bm{x}}_2 = \left[\begin{array}{c} 
	-1\\1\\0\\0\end{array}\right]
\end{equation}
\begin{equation}
	\vec{\bm{x}}_3 = \left[\begin{array}{c} 
	\frac{(1-p_m)R_m-R_g/2}{R_g-R_m}\\
	\frac{p_mR_m-R_g/2}{R_g-R_m}\\
	0\\
	1\\
	\end{array}\right]
\end{equation}
 and, for $\vec{\bm{x}}_4$,
 \begin{align}
 	\vec{\bm{x}}_4(1) &= \frac{(1-p_m)R_m}{R_g-R_o}-\frac{\frac{R_g}{2}R_m}{R_o(R_g-R_o)}\\ 
 	&\phantom{=}- \frac{p_o}{1-p_o}\frac{R_m-R_o}{R_o}\frac{\frac{R_g}{2}-(1-p_g)R_o}{R_g-R_o}\notag\\ 
 	 \vec{\bm{x}}_4(2) &=\frac{p_mR_m}{R_g-R_o}-\frac{\frac{R_g}{2}R_m}{R_o(R_g-R_o)}\\ 
 	&\phantom{=}- \frac{p_o}{1-p_o}\frac{R_m-R_o}{R_o}\frac{\frac{R_g}{2}-p_gR_o}{R_g-R_o}\notag\\  
 	\vec{\bm{x}}_4(3) &= \frac{R_m-R_o}{(1-p_o)R_o}\\
	 \vec{\bm{x}}_4(4) &= 1
 \end{align}

The constants $c$ are dependent on the initial populations $n_1(0),n_2(0),n_3(0),n_4(0)$:
\begin{widetext}
	\begin{align}
	c_1 &=\frac{n_1(0)+n_2(0)+n_3(0)+n_4(0)}{2} = \frac{1}{2}\\
	c_2 &=\frac{1}{2}(n_2(0)-n_1(0))+n_4(0)\left(\frac{1}{2}-p_m\right)\frac{R_m}{R_g-R_m} \\
	& \phantom{=}+n_3(0)\left[ p_o\left(\frac{1}{2}-p_g\right)\frac{R_o}{R_g-R_o}-(1-p_o)\left(\frac{1}{2}-p_m\right)\frac{R_m}{R_g-R_m}		\frac{R_o}{R_g-R_o}\right]\notag\\
	c_3 &= n_4(0)-n_3(0)(1-p_o)\frac{R_o}{R_m-R_o}\\
	c_4&= n_3(0)(1-p_o)\frac{R_o}{R_m-R_o}
	\end{align}
\end{widetext}

For a stimulated echo, the relevant initial populations $n_i(0)$ are the populations after the first two pulses. Half the population resides in the non-resonant hyperfine state, $n_1(0) = 0.5$, while the other half is divided between the second hyperfine state and the excited state, $n_2(0) = 0.5-d_0, n_3(0) = d_0$, where $d_0$ is a measure of the depth of the grating. The constants $c$ can, therefore, be slightly simplified:
\begin{align}
	 c_1 &= \frac{1}{2}\\
	c_2 &=d_0\left( -\frac{1}{2}+p_o\left(\frac{1}{2}-p_g\right)\frac{R_o}{R_g-R_o} \right. \\
	&\phantom{=d_0\left(\right.}\left.-(1-p_o)\left(\frac{1}{2}-p_m\right)\frac{R_m}{R_g-R_m}\frac{R_o}{R_g-R_o} \right)\notag\\
	c_3 &= -d_0(1-p_o)\frac{R_o}{R_m-R_o}\\
	c_4&= d_0(1-p_o)\frac{R_o}{R_m-R_o}
\end{align}

\section{Structure of the $\Delta_g -\Delta_e$ antihole}\label{appendixb}
In Section \ref{sec:zeeman}, we calculated the structure of the antihole at $\Delta_g$ due to the effects of the inhomogeneous broadening $G$, the resonance lineshape $k_{ah}$ and the hole lineshape $\zeta$ (Equation \eqref{eq:aholeamp}). This antihole was chosen because it gives a simple expression, but in the experiment it is most often the $\Delta_g -\Delta_e$ antihole that is seen. Because the shift of this antihole is not equal to the ground state splitting, the equation is more complicated. 

To derive an expression for the antihole at $\Delta_g -\Delta_e$, we first make the assumption that the ground and excited state splittings are completely correlated so that an ion with a ground state splitting $\nu_c$ will appear in the antihole at a frequency $\nu_c \epsilon$ where $\epsilon = (1-\frac{\Delta_e}{\Delta_g})$. This assumption can be justified by looking at the holeburning structure in Figure~\ref{fig:ext001}: the structure of the antihole at $\Delta_g-\Delta_e$ is identical to that at $\Delta_g$, indicating that the ground and excited state splittings are well correlated.  The lineshape for the antihole at $\Delta_g-\Delta_e$ is then
\begin{align}
	H_{ah}(f,f_c) = \int\limits_{-\infty}^{+\infty} &k_{ah}\left(\frac{\delta\nu+\alpha-\nu_{Al}}{\epsilon}\right)\notag \\&\times G(\delta\nu+\alpha-\Delta_g\epsilon)\zeta(\alpha)d\alpha
	\label{eq:hah}
\end{align}


\begin{thebibliography}{24}%
\makeatletter
\providecommand \@ifxundefined [1]{%
 \@ifx{#1\undefined}
}%
\providecommand \@ifnum [1]{%
 \ifnum #1\expandafter \@firstoftwo
 \else \expandafter \@secondoftwo
 \fi
}%
\providecommand \@ifx [1]{%
 \ifx #1\expandafter \@firstoftwo
 \else \expandafter \@secondoftwo
 \fi
}%
\providecommand \natexlab [1]{#1}%
\providecommand \enquote  [1]{``#1''}%
\providecommand \bibnamefont  [1]{#1}%
\providecommand \bibfnamefont [1]{#1}%
\providecommand \citenamefont [1]{#1}%
\providecommand \href@noop [0]{\@secondoftwo}%
\providecommand \href [0]{\begingroup \@sanitize@url \@href}%
\providecommand \@href[1]{\@@startlink{#1}\@@href}%
\providecommand \@@href[1]{\endgroup#1\@@endlink}%
\providecommand \@sanitize@url [0]{\catcode `\\12\catcode `\$12\catcode
  `\&12\catcode `\#12\catcode `\^12\catcode `\_12\catcode `\%12\relax}%
\providecommand \@@startlink[1]{}%
\providecommand \@@endlink[0]{}%
\providecommand \url  [0]{\begingroup\@sanitize@url \@url }%
\providecommand \@url [1]{\endgroup\@href {#1}{\urlprefix }}%
\providecommand \urlprefix  [0]{URL }%
\providecommand \Eprint [0]{\href }%
\providecommand \doibase [0]{http://dx.doi.org/}%
\providecommand \selectlanguage [0]{\@gobble}%
\providecommand \bibinfo  [0]{\@secondoftwo}%
\providecommand \bibfield  [0]{\@secondoftwo}%
\providecommand \translation [1]{[#1]}%
\providecommand \BibitemOpen [0]{}%
\providecommand \bibitemStop [0]{}%
\providecommand \bibitemNoStop [0]{.\EOS\space}%
\providecommand \EOS [0]{\spacefactor3000\relax}%
\providecommand \BibitemShut  [1]{\csname bibitem#1\endcsname}%
\let\auto@bib@innerbib\@empty
\bibitem [{\citenamefont {Babbitt}\ \emph {et~al.}(2014)\citenamefont
  {Babbitt}, \citenamefont {Barber}, \citenamefont {Bekker}, \citenamefont
  {Chase}, \citenamefont {Harrington}, \citenamefont {Merkel}, \citenamefont
  {Mohan}, \citenamefont {Sharpe}, \citenamefont {Stiffler}, \citenamefont
  {Traxinger},\ and\ \citenamefont {Woidtke}}]{babbitt14}%
  \BibitemOpen
  \bibfield  {author} {\bibinfo {author} {\bibfnamefont {W.~R.}\ \bibnamefont
  {Babbitt}}, \bibinfo {author} {\bibfnamefont {Z.~W.}\ \bibnamefont {Barber}},
  \bibinfo {author} {\bibfnamefont {S.~H.}\ \bibnamefont {Bekker}}, \bibinfo
  {author} {\bibfnamefont {M.~D.}\ \bibnamefont {Chase}}, \bibinfo {author}
  {\bibfnamefont {C.}~\bibnamefont {Harrington}}, \bibinfo {author}
  {\bibfnamefont {K.~D.}\ \bibnamefont {Merkel}}, \bibinfo {author}
  {\bibfnamefont {R.~K.}\ \bibnamefont {Mohan}}, \bibinfo {author}
  {\bibfnamefont {T.}~\bibnamefont {Sharpe}}, \bibinfo {author} {\bibfnamefont
  {C.~R.}\ \bibnamefont {Stiffler}}, \bibinfo {author} {\bibfnamefont {A.~S.}\
  \bibnamefont {Traxinger}}, \ and\ \bibinfo {author} {\bibfnamefont {A.~J.}\
  \bibnamefont {Woidtke}},\ }\href {\doibase 10.1088/1054-660X/24/9/094002}
  {\bibfield  {journal} {\bibinfo  {journal} {Laser Phys.}\ }\textbf {\bibinfo
  {volume} {24}},\ \bibinfo {pages} {094002} (\bibinfo {year}
  {2014})}\BibitemShut {NoStop}%
\bibitem [{\citenamefont {Thiel}\ \emph {et~al.}(2011)\citenamefont {Thiel},
  \citenamefont {B\"ottger},\ and\ \citenamefont {Cone}}]{thiel11}%
  \BibitemOpen
  \bibfield  {author} {\bibinfo {author} {\bibfnamefont {C.}~\bibnamefont
  {Thiel}}, \bibinfo {author} {\bibfnamefont {T.}~\bibnamefont {B\"ottger}}, \
  and\ \bibinfo {author} {\bibfnamefont {R.}~\bibnamefont {Cone}},\ }\href
  {\doibase 10.1016/j.jlumin.2010.12.015} {\bibfield  {journal} {\bibinfo
  {journal} {J. Lumin.}\ }\textbf {\bibinfo {volume} {131}},\ \bibinfo {pages}
  {353} (\bibinfo {year} {2011})}\BibitemShut {NoStop}%
\bibitem [{\citenamefont {Tittel}\ \emph {et~al.}(2010)\citenamefont {Tittel},
  \citenamefont {Afzelius}, \citenamefont {Chaneli\`ere}, \citenamefont {Cone},
  \citenamefont {Kr\"oll}, \citenamefont {Moiseev},\ and\ \citenamefont
  {Sellars}}]{tittel10}%
  \BibitemOpen
  \bibfield  {author} {\bibinfo {author} {\bibfnamefont {W.}~\bibnamefont
  {Tittel}}, \bibinfo {author} {\bibfnamefont {M.}~\bibnamefont {Afzelius}},
  \bibinfo {author} {\bibfnamefont {T.}~\bibnamefont {Chaneli\`ere}}, \bibinfo
  {author} {\bibfnamefont {R.}~\bibnamefont {Cone}}, \bibinfo {author}
  {\bibfnamefont {S.}~\bibnamefont {Kr\"oll}}, \bibinfo {author} {\bibfnamefont
  {S.~A.}\ \bibnamefont {Moiseev}}, \ and\ \bibinfo {author} {\bibfnamefont
  {M.}~\bibnamefont {Sellars}},\ }\href {\doibase 10.1002/lpor.200810056}
  {\bibfield  {journal} {\bibinfo  {journal} {Laser Photon. Rev.}\ }\textbf
  {\bibinfo {volume} {4}},\ \bibinfo {pages} {244} (\bibinfo {year}
  {2010})}\BibitemShut {NoStop}%
\bibitem [{\citenamefont {B\"ottger}\ \emph {et~al.}(2006)\citenamefont
  {B\"ottger}, \citenamefont {Thiel}, \citenamefont {Sun},\ and\ \citenamefont
  {Cone}}]{bottger06}%
  \BibitemOpen
  \bibfield  {author} {\bibinfo {author} {\bibfnamefont {T.}~\bibnamefont
  {B\"ottger}}, \bibinfo {author} {\bibfnamefont {C.~W.}\ \bibnamefont
  {Thiel}}, \bibinfo {author} {\bibfnamefont {Y.}~\bibnamefont {Sun}}, \ and\
  \bibinfo {author} {\bibfnamefont {R.~L.}\ \bibnamefont {Cone}},\ }\href
  {\doibase 10.1103/PhysRevB.73.075101} {\bibfield  {journal} {\bibinfo
  {journal} {Phys. Rev. B}\ }\textbf {\bibinfo {volume} {73}},\ \bibinfo
  {pages} {075101} (\bibinfo {year} {2006})}\BibitemShut {NoStop}%
\bibitem [{\citenamefont {Zhong}\ \emph {et~al.}(2015)\citenamefont {Zhong},
  \citenamefont {Hedges}, \citenamefont {Ahlefeldt}, \citenamefont
  {Bartholomew}, \citenamefont {Beavan}, \citenamefont {Wittig}, \citenamefont
  {Longdell},\ and\ \citenamefont {Sellars}}]{zhong15}%
  \BibitemOpen
  \bibfield  {author} {\bibinfo {author} {\bibfnamefont {M.}~\bibnamefont
  {Zhong}}, \bibinfo {author} {\bibfnamefont {M.~P.}\ \bibnamefont {Hedges}},
  \bibinfo {author} {\bibfnamefont {R.~L.}\ \bibnamefont {Ahlefeldt}}, \bibinfo
  {author} {\bibfnamefont {J.~G.}\ \bibnamefont {Bartholomew}}, \bibinfo
  {author} {\bibfnamefont {S.~E.}\ \bibnamefont {Beavan}}, \bibinfo {author}
  {\bibfnamefont {S.~M.}\ \bibnamefont {Wittig}}, \bibinfo {author}
  {\bibfnamefont {J.~J.}\ \bibnamefont {Longdell}}, \ and\ \bibinfo {author}
  {\bibfnamefont {M.~J.}\ \bibnamefont {Sellars}},\ }\href {\doibase
  10.1038/nature14025} {\bibfield  {journal} {\bibinfo  {journal} {Nature}\
  }\textbf {\bibinfo {volume} {517}},\ \bibinfo {pages} {177} (\bibinfo {year}
  {2015})}\BibitemShut {NoStop}%
\bibitem [{\citenamefont {Anderson}\ and\ \citenamefont
  {Freeman}(1962)}]{anderson62}%
  \BibitemOpen
  \bibfield  {author} {\bibinfo {author} {\bibfnamefont {W.~A.}\ \bibnamefont
  {Anderson}}\ and\ \bibinfo {author} {\bibfnamefont {R.}~\bibnamefont
  {Freeman}},\ }\href {\doibase 10.1063/1.1732980} {\bibfield  {journal}
  {\bibinfo  {journal} {J. Chem. Phys.}\ }\textbf {\bibinfo {volume} {37}},\
  \bibinfo {pages} {85} (\bibinfo {year} {1962})}\BibitemShut {NoStop}%
\bibitem [{\citenamefont {Hartmann}\ and\ \citenamefont
  {Hahn}(1962)}]{hartmann62}%
  \BibitemOpen
  \bibfield  {author} {\bibinfo {author} {\bibfnamefont {S.~R.}\ \bibnamefont
  {Hartmann}}\ and\ \bibinfo {author} {\bibfnamefont {E.~L.}\ \bibnamefont
  {Hahn}},\ }\href {\doibase 10.1103/PhysRev.128.2042} {\bibfield  {journal}
  {\bibinfo  {journal} {Phys. Rev.}\ }\textbf {\bibinfo {volume} {128}},\
  \bibinfo {pages} {2042} (\bibinfo {year} {1962})}\BibitemShut {NoStop}%
\bibitem [{\citenamefont {Anderson}\ and\ \citenamefont
  {Hartmann}(1962)}]{anderson62a}%
  \BibitemOpen
  \bibfield  {author} {\bibinfo {author} {\bibfnamefont {A.~G.}\ \bibnamefont
  {Anderson}}\ and\ \bibinfo {author} {\bibfnamefont {S.~R.}\ \bibnamefont
  {Hartmann}},\ }\href {\doibase 10.1103/PhysRev.128.2023} {\bibfield
  {journal} {\bibinfo  {journal} {Phys. Rev.}\ }\textbf {\bibinfo {volume}
  {128}},\ \bibinfo {pages} {2023} (\bibinfo {year} {1962})}\BibitemShut
  {NoStop}%
\bibitem [{\citenamefont {Wald}\ \emph {et~al.}(1992)\citenamefont {Wald},
  \citenamefont {Lukac},\ and\ \citenamefont {Hahn}}]{wald92}%
  \BibitemOpen
  \bibfield  {author} {\bibinfo {author} {\bibfnamefont {L.~L.}\ \bibnamefont
  {Wald}}, \bibinfo {author} {\bibfnamefont {M.}~\bibnamefont {Lukac}}, \ and\
  \bibinfo {author} {\bibfnamefont {E.~L.}\ \bibnamefont {Hahn}},\ }\href
  {\doibase 10.1364/JOSAB.9.000789} {\bibfield  {journal} {\bibinfo  {journal}
  {J. Opt. Soc. Am. B}\ }\textbf {\bibinfo {volume} {9}},\ \bibinfo {pages}
  {789} (\bibinfo {year} {1992})}\BibitemShut {NoStop}%
\bibitem [{\citenamefont {Otto}\ \emph {et~al.}(1986)\citenamefont {Otto},
  \citenamefont {Lukac},\ and\ \citenamefont {Hahn}}]{otto86}%
  \BibitemOpen
  \bibfield  {author} {\bibinfo {author} {\bibfnamefont {F.~W.}\ \bibnamefont
  {Otto}}, \bibinfo {author} {\bibfnamefont {M.}~\bibnamefont {Lukac}}, \ and\
  \bibinfo {author} {\bibfnamefont {E.~L.}\ \bibnamefont {Hahn}},\ }\href
  {\doibase 10.1016/0022-2313(86)90018-9} {\bibfield  {journal} {\bibinfo
  {journal} {J. Lumin.}\ }\textbf {\bibinfo {volume} {35}},\ \bibinfo {pages}
  {321} (\bibinfo {year} {1986})}\BibitemShut {NoStop}%
\bibitem [{\citenamefont {Lukac}\ \emph {et~al.}(1989)\citenamefont {Lukac},
  \citenamefont {Otto},\ and\ \citenamefont {Hahn}}]{lukac89}%
  \BibitemOpen
  \bibfield  {author} {\bibinfo {author} {\bibfnamefont {M.}~\bibnamefont
  {Lukac}}, \bibinfo {author} {\bibfnamefont {F.~W.}\ \bibnamefont {Otto}}, \
  and\ \bibinfo {author} {\bibfnamefont {E.~L.}\ \bibnamefont {Hahn}},\ }\href
  {\doibase 10.1103/PhysRevA.39.1123} {\bibfield  {journal} {\bibinfo
  {journal} {Phys. Rev. A}\ }\textbf {\bibinfo {volume} {39}},\ \bibinfo
  {pages} {1123} (\bibinfo {year} {1989})}\BibitemShut {NoStop}%
\bibitem [{\citenamefont {Macfarlane}(1993)}]{macfarlane93}%
  \BibitemOpen
  \bibfield  {author} {\bibinfo {author} {\bibfnamefont {R.~M.}\ \bibnamefont
  {Macfarlane}},\ }\href {\doibase 10.1364/OL.18.001958} {\bibfield  {journal}
  {\bibinfo  {journal} {Opt. Lett.}\ }\textbf {\bibinfo {volume} {18}},\
  \bibinfo {pages} {1958} (\bibinfo {year} {1993})}\BibitemShut {NoStop}%
\bibitem [{\citenamefont {Guillot-No\"el}\ \emph {et~al.}(2005)\citenamefont
  {Guillot-No\"el}, \citenamefont {Goldner}, \citenamefont {Antic-Fidancev},\
  and\ \citenamefont {Le~Gou\"et}}]{guillot-noel05}%
  \BibitemOpen
  \bibfield  {author} {\bibinfo {author} {\bibfnamefont {O.}~\bibnamefont
  {Guillot-No\"el}}, \bibinfo {author} {\bibfnamefont {P.}~\bibnamefont
  {Goldner}}, \bibinfo {author} {\bibfnamefont {E.}~\bibnamefont
  {Antic-Fidancev}}, \ and\ \bibinfo {author} {\bibfnamefont {J.-L.}\
  \bibnamefont {Le~Gou\"et}},\ }\href {\doibase 10.1103/PhysRevB.71.174409}
  {\bibfield  {journal} {\bibinfo  {journal} {Phys. Rev. B}\ }\textbf {\bibinfo
  {volume} {71}},\ \bibinfo {pages} {174409} (\bibinfo {year}
  {2005})}\BibitemShut {NoStop}%
\bibitem [{\citenamefont {Heshami}\ \emph {et~al.}(2012)\citenamefont
  {Heshami}, \citenamefont {Green}, \citenamefont {Han}, \citenamefont {Rispe},
  \citenamefont {Saglamyurek}, \citenamefont {Sinclair}, \citenamefont
  {Tittel},\ and\ \citenamefont {Simon}}]{heshami12}%
  \BibitemOpen
  \bibfield  {author} {\bibinfo {author} {\bibfnamefont {K.}~\bibnamefont
  {Heshami}}, \bibinfo {author} {\bibfnamefont {A.}~\bibnamefont {Green}},
  \bibinfo {author} {\bibfnamefont {Y.}~\bibnamefont {Han}}, \bibinfo {author}
  {\bibfnamefont {A.}~\bibnamefont {Rispe}}, \bibinfo {author} {\bibfnamefont
  {E.}~\bibnamefont {Saglamyurek}}, \bibinfo {author} {\bibfnamefont
  {N.}~\bibnamefont {Sinclair}}, \bibinfo {author} {\bibfnamefont
  {W.}~\bibnamefont {Tittel}}, \ and\ \bibinfo {author} {\bibfnamefont
  {C.}~\bibnamefont {Simon}},\ }\href {\doibase 10.1103/PhysRevA.86.013813}
  {\bibfield  {journal} {\bibinfo  {journal} {Phys. Rev. A}\ }\textbf {\bibinfo
  {volume} {86}},\ \bibinfo {pages} {013813} (\bibinfo {year}
  {2012})}\BibitemShut {NoStop}%
\bibitem [{\citenamefont {Trontelj}\ and\ \citenamefont
  {Schmidt}(1973)}]{trontelj73}%
  \BibitemOpen
  \bibfield  {author} {\bibinfo {author} {\bibfnamefont {Z.}~\bibnamefont
  {Trontelj}}\ and\ \bibinfo {author} {\bibfnamefont {V.~H.}\ \bibnamefont
  {Schmidt}},\ }\href {\doibase 10.1103/PhysRevB.7.4145} {\bibfield  {journal}
  {\bibinfo  {journal} {Phys. Rev. B}\ }\textbf {\bibinfo {volume} {7}},\
  \bibinfo {pages} {4145} (\bibinfo {year} {1973})}\BibitemShut {NoStop}%
\bibitem [{\citenamefont {de~Seze}\ \emph {et~al.}(2006)\citenamefont
  {de~Seze}, \citenamefont {Louchet}, \citenamefont {Crozatier}, \citenamefont
  {Lorger\'e}, \citenamefont {Bretenaker}, \citenamefont {Le~Gou\"et},
  \citenamefont {Guillot-No\"el},\ and\ \citenamefont {Goldner}}]{deseze06}%
  \BibitemOpen
  \bibfield  {author} {\bibinfo {author} {\bibfnamefont {F.}~\bibnamefont
  {de~Seze}}, \bibinfo {author} {\bibfnamefont {A.}~\bibnamefont {Louchet}},
  \bibinfo {author} {\bibfnamefont {V.}~\bibnamefont {Crozatier}}, \bibinfo
  {author} {\bibfnamefont {I.}~\bibnamefont {Lorger\'e}}, \bibinfo {author}
  {\bibfnamefont {F.}~\bibnamefont {Bretenaker}}, \bibinfo {author}
  {\bibfnamefont {J.-L.}\ \bibnamefont {Le~Gou\"et}}, \bibinfo {author}
  {\bibfnamefont {O.}~\bibnamefont {Guillot-No\"el}}, \ and\ \bibinfo {author}
  {\bibfnamefont {P.}~\bibnamefont {Goldner}},\ }\href {\doibase
  10.1103/PhysRevB.73.085112} {\bibfield  {journal} {\bibinfo  {journal}
  {Physical Review B}\ }\textbf {\bibinfo {volume} {73}},\ \bibinfo {pages}
  {085112} (\bibinfo {year} {2006})}\BibitemShut {NoStop}%
\bibitem [{\citenamefont {Louchet}\ \emph {et~al.}(2007)\citenamefont
  {Louchet}, \citenamefont {Habib}, \citenamefont {Crozatier}, \citenamefont
  {Lorger\'e}, \citenamefont {Goldfarb}, \citenamefont {Bretenaker},
  \citenamefont {{Le Gou\"et}}, \citenamefont {Guillot-No\"el},\ and\
  \citenamefont {Goldner}}]{louchet07}%
  \BibitemOpen
  \bibfield  {author} {\bibinfo {author} {\bibfnamefont {A.}~\bibnamefont
  {Louchet}}, \bibinfo {author} {\bibfnamefont {J.~S.}\ \bibnamefont {Habib}},
  \bibinfo {author} {\bibfnamefont {V.}~\bibnamefont {Crozatier}}, \bibinfo
  {author} {\bibfnamefont {I.}~\bibnamefont {Lorger\'e}}, \bibinfo {author}
  {\bibfnamefont {F.}~\bibnamefont {Goldfarb}}, \bibinfo {author}
  {\bibfnamefont {F.}~\bibnamefont {Bretenaker}}, \bibinfo {author}
  {\bibfnamefont {J.-L.}\ \bibnamefont {{Le Gou\"et}}}, \bibinfo {author}
  {\bibfnamefont {O.}~\bibnamefont {Guillot-No\"el}}, \ and\ \bibinfo {author}
  {\bibfnamefont {P.}~\bibnamefont {Goldner}},\ }\href {\doibase
  10.1103/PhysRevB.75.035131} {\bibfield  {journal} {\bibinfo  {journal} {Phys.
  Rev. B}\ }\textbf {\bibinfo {volume} {75}},\ \bibinfo {pages} {035131}
  (\bibinfo {year} {2007})}\BibitemShut {NoStop}%
\bibitem [{\citenamefont {Sun}\ \emph {et~al.}(2000)\citenamefont {Sun},
  \citenamefont {Wang}, \citenamefont {Cone}, \citenamefont {Equall},\ and\
  \citenamefont {Leask}}]{sun00}%
  \BibitemOpen
  \bibfield  {author} {\bibinfo {author} {\bibfnamefont {Y.}~\bibnamefont
  {Sun}}, \bibinfo {author} {\bibfnamefont {G.~M.}\ \bibnamefont {Wang}},
  \bibinfo {author} {\bibfnamefont {R.~L.}\ \bibnamefont {Cone}}, \bibinfo
  {author} {\bibfnamefont {R.~W.}\ \bibnamefont {Equall}}, \ and\ \bibinfo
  {author} {\bibfnamefont {M.~J.~M.}\ \bibnamefont {Leask}},\ }\href {\doibase
  10.1103/PhysRevB.62.15443} {\bibfield  {journal} {\bibinfo  {journal} {Phys.
  Rev. B}\ }\textbf {\bibinfo {volume} {62}},\ \bibinfo {pages} {15443}
  (\bibinfo {year} {2000})}\BibitemShut {NoStop}%
\bibitem [{\citenamefont {Brog}\ \emph {et~al.}(1966)\citenamefont {Brog},
  \citenamefont {Jones~Jr.},\ and\ \citenamefont {Verber}}]{brog66}%
  \BibitemOpen
  \bibfield  {author} {\bibinfo {author} {\bibfnamefont {K.~C.}\ \bibnamefont
  {Brog}}, \bibinfo {author} {\bibfnamefont {W.~H.}\ \bibnamefont {Jones~Jr.}},
  \ and\ \bibinfo {author} {\bibfnamefont {C.~M.}\ \bibnamefont {Verber}},\
  }\href {\doibase 10.1016/0031-9163(66)90353-2} {\bibfield  {journal}
  {\bibinfo  {journal} {Phys. Lett.}\ }\textbf {\bibinfo {volume} {20}},\
  \bibinfo {pages} {258} (\bibinfo {year} {1966})}\BibitemShut {NoStop}%
\bibitem [{\citenamefont {Caird}\ \emph {et~al.}(1975)\citenamefont {Caird},
  \citenamefont {DeShazer},\ and\ \citenamefont {Nella}}]{caird75}%
  \BibitemOpen
  \bibfield  {author} {\bibinfo {author} {\bibfnamefont {J.}~\bibnamefont
  {Caird}}, \bibinfo {author} {\bibfnamefont {L.}~\bibnamefont {DeShazer}}, \
  and\ \bibinfo {author} {\bibfnamefont {J.}~\bibnamefont {Nella}},\ }\href
  {\doibase 10.1109/JQE.1975.1068541} {\bibfield  {journal} {\bibinfo
  {journal} {{IEEE} J. Quantum Elect.}\ }\textbf {\bibinfo {volume} {11}},\
  \bibinfo {pages} {874} (\bibinfo {year} {1975})}\BibitemShut {NoStop}%
\bibitem [{\citenamefont {Strickland}\ \emph {et~al.}(2000)\citenamefont
  {Strickland}, \citenamefont {Sellin}, \citenamefont {Sun}, \citenamefont
  {Carlsten},\ and\ \citenamefont {Cone}}]{strickland00}%
  \BibitemOpen
  \bibfield  {author} {\bibinfo {author} {\bibfnamefont {N.~M.}\ \bibnamefont
  {Strickland}}, \bibinfo {author} {\bibfnamefont {P.~B.}\ \bibnamefont
  {Sellin}}, \bibinfo {author} {\bibfnamefont {Y.}~\bibnamefont {Sun}},
  \bibinfo {author} {\bibfnamefont {J.~L.}\ \bibnamefont {Carlsten}}, \ and\
  \bibinfo {author} {\bibfnamefont {R.~L.}\ \bibnamefont {Cone}},\ }\href
  {\doibase 10.1103/PhysRevB.62.1473} {\bibfield  {journal} {\bibinfo
  {journal} {Phys. Rev. B}\ }\textbf {\bibinfo {volume} {62}},\ \bibinfo
  {pages} {1473} (\bibinfo {year} {2000})}\BibitemShut {NoStop}%
\bibitem [{\citenamefont {Alff}\ and\ \citenamefont {Wertheim}(1961)}]{alff61}%
  \BibitemOpen
  \bibfield  {author} {\bibinfo {author} {\bibfnamefont {C.}~\bibnamefont
  {Alff}}\ and\ \bibinfo {author} {\bibfnamefont {G.~K.}\ \bibnamefont
  {Wertheim}},\ }\href {\doibase 10.1103/PhysRev.122.1414} {\bibfield
  {journal} {\bibinfo  {journal} {Phys. Rev.}\ }\textbf {\bibinfo {volume}
  {122}},\ \bibinfo {pages} {1414} (\bibinfo {year} {1961})}\BibitemShut
  {NoStop}%
\bibitem [{\citenamefont {Euler}\ and\ \citenamefont {Bruce}(1965)}]{euler65}%
  \BibitemOpen
  \bibfield  {author} {\bibinfo {author} {\bibfnamefont {F.}~\bibnamefont
  {Euler}}\ and\ \bibinfo {author} {\bibfnamefont {J.~A.}\ \bibnamefont
  {Bruce}},\ }\href {\doibase 10.1107/S0365110X65004747} {\bibfield  {journal}
  {\bibinfo  {journal} {Acta Crystallogr.}\ }\textbf {\bibinfo {volume} {19}},\
  \bibinfo {pages} {971} (\bibinfo {year} {1965})}\BibitemShut {NoStop}%
\bibitem [{\citenamefont {Shelyapina}\ \emph {et~al.}(2006)\citenamefont
  {Shelyapina}, \citenamefont {Kasperovich},\ and\ \citenamefont
  {Wolfers}}]{shelyapina06}%
  \BibitemOpen
  \bibfield  {author} {\bibinfo {author} {\bibfnamefont {M.~G.}\ \bibnamefont
  {Shelyapina}}, \bibinfo {author} {\bibfnamefont {V.~S.}\ \bibnamefont
  {Kasperovich}}, \ and\ \bibinfo {author} {\bibfnamefont {P.}~\bibnamefont
  {Wolfers}},\ }\href {\doibase 10.1016/j.jpcs.2005.10.181} {\bibfield
  {journal} {\bibinfo  {journal} {J. Phys. Chem. Solids}\ }\textbf {\bibinfo
  {volume} {67}},\ \bibinfo {pages} {720} (\bibinfo {year} {2006})}\BibitemShut
  {NoStop}%
\end{thebibliography}

%

\end{document}